\documentclass[a4paper,11pt]{article}
\pdfoutput=1

\usepackage{jheppub1}
\usepackage[T1]{fontenc}
\usepackage{amssymb}
\usepackage{graphicx}
\usepackage{amsmath}
\usepackage{hyperref}
\usepackage{tensor}
\usepackage{epstopdf}
\usepackage{extarrows}
\usepackage{verbatim}
\usepackage{subfigure}
\usepackage{float}
\usepackage{tikz}

\newcommand{\td}{\text{d}}

\begin{document}
	
	\title{Charged Taub-NUT type Black Holes in Einstein-Weyl-Maxwell Theory}
	\renewcommand {\thefootnote} {\fnsymbol {footnote}}
    \author{Ze Li and Hai-Shan Liu \footnote{Corresponding author}  }
    \emailAdd{lize@tju.edu.cn}
    \emailAdd{hsliu.zju@gmail.com  }
    \affiliation{Center for Joint Quantum Studies and Department of Physics, School of Science, Tianjin University, Yaguan Road 135, Jinnan District, 300350 Tianjin, P.~R.~China }
	
	\abstract{

		We numerically construct new charged Taub-NUT-type black hole solutions in four-dimensional Einstein-Weyl-Maxwell theory. We explore the effects of the NUT parameter and the electric charge parameter on the black hole solutions in great detail. Compared with static spherical black holes in the same theory, there is one major difference: there are three branches of NUT solutions, whereas there are only two branches of static spherical black hole solutions. Compared with neutral NUT solutions in Weyl gravity, where the solutions intersect with each other, the charged NUT solutions we find are disconnected.
		
	}

	\maketitle
	%\tableofcontents
	 \renewcommand {\thefootnote} {\arabic {footnote}}
	
\section{Introduction}
Since its inception, general relativity has withstood numerous experimental and observational tests, achieving remarkable success. In the 21st century, the direct detection of gravitational waves and imaging of black holes have further confirmed the validity of Einstein's theory in strong gravitational fields~\cite{LIGOScientific:2017vwq,LIGOScientific:2017zic,EventHorizonTelescope:2019dse,EventHorizonTelescope:2019uob,EventHorizonTelescope:2019jan,EventHorizonTelescope:2019ths,EventHorizonTelescope:2019pgp,EventHorizonTelescope:2019ggy}. However, many pieces of evidence suggest that general relativity is not the ultimate theory of gravity. Experimentally, it fails to naturally explain phenomena such as the accelerated expansion of the universe. General relativity exhibits non-renormalizability during quantization, making it incompatible with quantum field theory.

To address these challenges, various modified and extended theories of gravity have been proposed. Studies show that theories of gravity with quadratic curvature terms exhibit renormalizability, but typically introduce higher-derivative terms that may lead to ghost modes~\cite{Stelle:1976gc}, thereby compromising theoretical stability. To circumvent this issue, researchers have developed several approaches including chiral gravity~\cite{Li:2008dq}, topological gravity~\cite{Deser:1981wh}, and new massive gravity~\cite{Bergshoeff:2009aq} in three dimensions. Subsequently, Lu and Pope proposed the four-dimensional critical gravity theory, demonstrating that under specific parameters, Einstein gravity extended with quadratic curvature terms can avoid ghost modes~\cite{Lu:2011zk}. This theoretical framework has since been generalized to higher dimensions~\cite{Deser:2011xc}.

Black hole solutions are crucial for exploring various properties of gravity, such as Schwarzschild and Reissner-Nordström (RN) black holes. However, constructing black hole solutions in higher-derivative gravity theories remains a challenging task. No static spherically symmetric black hole solutions beyond the Schwarzschild metric had been found in Einstein gravity extended with quadratic curvature terms until 2015, when Lu et al. successfully constructed a new solution~\cite{Lu:2015cqa}. Additional static spherically symmetric solutions were later discovered in Einstein gravity plus quadratic curvature terms~\cite{ Huang:2022urr}.

The generalization to include charge has been further investigated in two contexts:

Electrically charged black holes by coupling Maxwell fields to higher-derivative theories~\cite{Lin:2016jjl,Wu:2019uvq};
Black holes with NUT charge in pure higher-derivative gravity (without matter fields)~\cite{Chen:2024hsh}.

In this work, we unify these approaches by constructing black hole solutions carrying both NUT and electric charges in higher-derivative gravity theories with Maxwell fields. We further analyze the combined effects of NUT and electric charges on black hole properties.

The structure of this paper is organized as follows. In Section \ref{section 2}, we introduce the Einstein-Weyl-Maxwell theory and analyze the characteristics of black hole solutions in this theory. In Section \ref{section 3}, we outline our parameter settings and use numerical method to construct black hole solutions. We separately fix the NUT parameter $n$ and the coupling parameter $a_1$, conducting a detailed examination of black hole properties within this framework.
Section \ref{section 4} summarizes our findings, highlighting the remarkable features of Taub-NUT type black holes in Einstein-Weyl-Maxwell theory.

\section{Einstein-Weyl-Maxwell theory and field equations}\label{section 2}
We consider the Einstein-Weyl-Maxwell theory in four-dimensional spacetime, with the Lagrangian given by
\begin{equation}
	\label{eq20012}
	\mathcal{L}=R-\alpha C_{\mu \nu \rho \sigma}C^{\mu \nu \rho \sigma}  - F_{\mu \nu} F^{\mu \nu}.
\end{equation}
where $\alpha$ is a coupling constants, $C_{\mu \nu \rho \sigma}$ is the Weyl tensor, $F_{\mu \nu} = \nabla_{\mu} A_{\nu}-\nabla_{\nu} A_{\nu}$ is the electromagnetic field strength tensor.   The gravitational part of the theory contains two modes: a massless spin-2 graviton (the usual Einstein mode), and a  massive spin-2 graviton with mass $m_{2}=1/\sqrt{2\alpha}$. The massive graviton contributes a form of  Yukawa-type potential ($\frac{1}{r}\text{e}^{\pm m_{2}r}$) in the metric functions in at large $r$ ~\cite{Smilga_2014,Lu:2015cqa,L__2017}.  A key feature of this theory is the decoupling of the scalar mode: for static, spherically symmetric black holes, the Ricci scalar $R$ vanishes identically.

It is well-known that the trace of the energy-momentum tensor of the Maxwell field vanishes in four dimensions, which implies that the Maxwell field does not source the trace part of the Einstein equations. Consequently, the vanishing Ricci scalar ($R=0$) remains unaffected by the Maxwell field.

By separately varying the metric $g_{\mu\nu}$ and the gauge field $A_{\mu}$, we obtain the equations of motion:
\begin{equation}
	\label{eq2002}
	\begin{aligned}
		R_{\mu \nu}-\frac{1}{2} g_{\mu \nu} R-4 \alpha B_{\mu \nu}-8 \pi T_{\mu \nu} & =0, \\
		\nabla_\mu F^{\mu \nu} & =0 .
	\end{aligned}
\end{equation}
where $B_{\mu \nu}$ is the Bach tensor, defined as
\begin{equation}
	\label{eq2003}
	B_{\mu \nu}=\left(\nabla^\rho \nabla^\sigma+\frac{1}{2} R^{\rho \sigma}\right) C_{\mu \rho \nu \sigma},
\end{equation}
and $T_{\mu \nu}$ is the energy-momentum tensor of the Maxwell field
\begin{equation}
	\label{eq2004}
	T_{\mu \nu}=F_{\mu \alpha} F_\nu{ }^\alpha-\frac{1}{4} g_{\mu \nu} F_{\rho \sigma} F^{\rho \sigma} .
\end{equation}
Both $T_{\mu\nu}$ and $B_{\mu\nu}$ are traceless, and the trace of the Einstein equations directly implies $R=0$, consistent with the initial statement in this section.

The Plebanski solution, a cohomogeneity-two metric, is a significant solution of Einstein-Maxwell gravity~\cite{Plebanski:1976gy}. The thermodynamics of this solution was studied recently by treating NUT parameter as an independent charge~\cite{Liu:2022wku}. The solution with both NUT and electric charges has the form
\begin{equation}
\text{d}s^2 = - f_0 \text{d}t^2 + \frac{\text{d}r^2}{f_0} + (r^2+n^2) (\text{d}\theta^2+\sin^2 \theta \text{d}\phi^2) \,,
\end{equation}
with
\begin{equation}
f_0 = \frac{r^2- 2 m r - n^2}{r^2+n^2} \,.
\end{equation}
And the Maxwell field is given by
  \begin{equation}
  A =  \frac{e r}{r^2+n^2} (\text{d}t + 2 n \cos \theta \text{d} \phi) \,.
  \end{equation}
The solution involves three integration constants: $m, n$, and $e$. However, when higher-curvature terms are introduced, the theory no longer admits the above NUT solution. Inspired by the construction of numerical charged solutions in higher-derivative gravity theories, we aim to construct new charged NUT solutions in Einstein-Weyl-Maxwell theory. The ansatz we consider is:
\begin{equation}
	\label{eq2005}
	\begin{aligned}
		\td s^{2}&=-h(r)  \left(\td t + 2n\cos\theta \td \phi \right) ^{2}+\frac{\td r^{2}}{f(r)}+(r^{2}+n^{2}) (\td \theta^2 +\sin^2\theta \td \phi^2), \\
		A &= - a (r)\left(\td t +2n \cos \theta \td \phi\right),
	\end{aligned}
\end{equation}
where $n$ is the NUT parameter. The metric functions $h(r)$, $f(r)$ and electric potential $a(r)$ are to be determined.
 
Substituting the ansatz \eqref{eq2005} into equations of motion \eqref{eq2002}, we obtain three independent equations:
\begin{equation}
	\label{eq2006}
	\begin{aligned}
	&h''+\frac{h'}{2}\left(\frac{f'}{f}-\frac{h'}{h}\right)+\frac{2}{r^{2}+n^{2}}\left[\left(r h\right)'+\frac{h}{f}\left(r f'-1\right)\right]+\frac{2 n^2}{(r^{2}+n^{2})^2}\frac{h}{f}\left(f-h\right)=0,\\
	&f''+\frac{1}{2 \alpha r \left(\left( r^2+n^2\right) h'- 2 r h\right)}\left\{-2 \left( r^2+n^2\right)^2 a^2  \right.\\
	&\left. +\left( r^2+n^2\right)\left[2\left(\frac{h}{f}-r f'\right)+\alpha r h'^{2}\left(\frac{f}{h}\right)^2\left(\frac{h}{f}\right)' \right] +\frac{2h}{f} \left(r^2 f +n^2 h +4 n^2 a^2\right)  \right.\\
	&\left. + \alpha \left[8\left(h+r h'\left(r^2 f' +2 n^2 f'-4 n^2 h'\right)\right) +\frac{4 r h^2 f'-3 r^2 h^2 f'^2 +2 n^2 f^2 h'^2 +3r^2 f^2 h'^2}{hf} \right]\right.\\
	&\left. +\frac{4 \alpha}{r^2+n^2}\left[r^3 h f' -2 n^2 r h' \left(f-4 h\right)- \frac{n^2 h^2 \left(r f' +2\right)}{f}\right]\right.\\
	&\left. +\frac{4 \alpha}{(r^2+n^2)^2}\frac{h}{f}\left[ \left(2 r^4 +8 r^2 n^2+3 n^4\right)f^2 -2n^2 f h\left(5r^2+n^2\right)-5n^2 h^2\right] \right\}=0,\\
	&a''+\frac{1}{2} a' \left(\frac{f'}{f}-\frac{h'}{h}+\frac{4 r}{ r^2+n^2}\right)+a \frac{4 n^2}{ \left( r^2+n^2\right)^2}\frac{h}{f}=0.
	\end{aligned}
\end{equation}
Due to the vanishing of Ricci scalar $R=0$, the three differential equations for $ h(r) $, $ f(r) $, and $ a(r) $ are of second order. Even with these simple second-order differential equations, constructing an analytic black hole solution is challenging. Therefore, we turn to numerical methods.

Before performing numerical calculations, we analyze the behavior of the metric functions near the black hole horizon and at infinity. We perform a Taylor expansion of $h(r), f(r)$ and $a(r)$ around the horizon radius $r_{0}$, which takes the form 
\begin{equation}
	\label{eq2007}
	\begin{aligned}
		&h(r)=h_1(r-r_{0})+h_2(r-r_{0})^2+h_3(r-r_{0})^3+\dots \,,\\
		&f(r)=f_1(r-r_{0})+f_2(r-r_{0})^2+f_3(r-r_{0})^3+\dots \,,\\
		&a(r)=a_0+a_1(r-r_{0})+a_2(r-r_{0})^2+a_3(r-r_{0})^3+\dots \,.
	\end{aligned}
\end{equation}
Substituting the Taylor expansions \eqref{eq2007} into equation \eqref{eq2006}, we find that  the high-order expansion coefficients can be impressed in terms of  $h_1, f_1, a_0, a_1$. For example, the second-order expansion coefficients are
\begin{equation}
	\label{eq2008}
	\begin{aligned}
		h_2=&\frac{h_1 \left(r_{0}^2+n^2-4 a_0^2   n^2+4 \alpha  f_1^2 \left(n^2-4 r_{0}^2\right)\right)}{8 \alpha  f_1^2 r_{0} \left(r_{0}^2+n^2\right)}\\
		&-\frac{h_1 \left(\left(4 \alpha  h_1 n^2+r_{0} \left( r_{0}^2+n^2-8 \alpha\right)\right)\right)+a_1^2   \left(r_{0}^2+n^2\right){}^2}{8 \alpha  f_1 r_{0} \left(r_{0}^2+n^2\right)},\\
		f_2=&\frac{12 a_0^2   n^2-4 \alpha  f_1^2 \left(4 r_{0}^2+3 n^2\right)-3 \left(r_{0}^2+n^2\right)}{8 \alpha  f_1 r_{0} \left(r_{0}^2+n^2\right)}\\
		&+\frac{3 a_1^2   \left(r_{0}^2+n^2\right){}^2+h_1 \left(12 \alpha  h_1 n^2+r_{0} \left(8 \alpha +3 r_{0}^2+3 n^2\right)\right)}{8 \alpha  h_1 r_{0} \left(r_{0}^2+n^2\right)},\\
		a_2=&-\frac{16 \alpha  a_0 h_1^2 n^2 r_{0}+a_1^3  \left(r_{0}^2+n^2\right){}^3}{8 \alpha  f_1 h_1 r_{0} \left(r_{0}^2+n^2\right){}^2}\\
		&-\frac{a_1 \left(n^2 \left(4 a_0^2  -4 \alpha  f_1^2+f_1 \left(4 \alpha  h_1+r_{0}\right)-1\right)+r_{0}^2 \left(8 \alpha  f_1^2+f_1 r_{0}-1\right)\right)}{8 \alpha  f_1^2 r_{0} \left(r_{0}^2+n^2\right)}.
	\end{aligned}
\end{equation}

Since the black hole entropy and temperature are closely related to the horizon, we can obtain these two thermodynamic quantities through the Talor expansion near the horizon. The temperature of the black hole can be calculated by the surface gravity $\kappa$
\begin{equation}
	\label{eq2015}
	T=\frac{\kappa}{2\pi}=\frac{\sqrt{h^{\prime}(r_{0})f^{\prime}(r_{0})}}{4\pi} =\frac{\sqrt{h_1f_1}}{4 \pi}.
\end{equation}
The entropy of the black hole can be calculated using the Wald entropy formula  \cite{Wald1983,IyerWald1994} 
\begin{equation}
	\label{eq2016}
	S=-\frac{1}{8}\int_{\Sigma} \frac{\delta\mathcal{L}}{\delta R_{\mu\nu\rho\sigma}} \epsilon_{\mu\nu}\epsilon_{\rho\sigma}\td \Sigma,
\end{equation}
where $\Sigma$ is an arbitrary cross-section of the Killing horizon. Considering the metric \eqref{eq2005}, $\td \Sigma$ can be expressed as $\td \Sigma=(r_{0}^{2}+n^2)\sin \theta \td \theta \text{d}\phi$. The $\epsilon_{\mu\nu}$ is binormal vector to the co-dimension 2 hypersurface $\Sigma$, which satisfies $\nabla_{a}\xi_{b}=\kappa\epsilon_{ab}$ with $\xi^{a}$ being the null Killing vector  and $\kappa$ being the surface gravity. With the coordinates $\{t,r,\theta,\phi\}$ and the ansatz \eqref{eq2005}, We can write $\epsilon_{\mu\nu}$ in components
\begin{equation}
	\label{eq201602}
	\epsilon_{\mu\nu}=\left(
	\begin{matrix}
		0 & \sqrt{h(r)/f(r)} & 0 & 0 \\
		-\sqrt{h(r)/f(r)} & 0 & 0 & -2n\cos\theta\sqrt{h(r)/f(r)} \\
		0 & 0 & 0 & 0 \\
		0 & 2n\cos\theta\sqrt{h(r)/f(r)} & 0 & 0 
	\end{matrix}
	\right).
\end{equation}
One can easily verify that $\epsilon_{\mu\nu}\epsilon^{\mu\nu}=-2$.
Through the Wald entropy formula, we observe that only the terms involving $R$ and $\alpha C_{\mu \nu \rho \sigma}C^{\mu \nu \rho \sigma}$ contribute to $\delta\mathcal{L}/\delta R_{\mu\nu\rho\sigma}$, thus the entropy is given by
\begin{equation}
	\label{eq2017}
		S=-\frac{1}{8}\int_{\Sigma} \left[\frac{1}{2}\left(g^{\mu \rho}g^{\nu \sigma}-g^{\mu \rho}g^{\nu \sigma}\right)-2\alpha C^{\mu\nu \rho\sigma}\right] \epsilon_{\mu\nu}\epsilon_{\rho\sigma}\td \Sigma.
\end{equation}
Evaluating this on the horizon yields
\begin{equation}
	\label{eq201702}
		S=\pi(r_{0}^2+n^2)\left(1-\frac{\alpha}{3}\left(\frac{4(1+r_{0}f^{\prime}(r_{0}))}{r_{0}^2+n^2}-\frac{3f^{\prime}(r_{0})}{2h^{\prime}(r_{0})}h^{\prime\prime}(r_{0})-\frac{1}{2}f^{\prime\prime}(r_{0})\right)\right).
\end{equation}
Using the expansion coefficients from \eqref{eq2008}, the entropy simplifies to
\begin{equation}
	\label{eq201703}
	S= \pi(r_{0}^2+n^2)-4\pi \alpha r_{0}f_1 \,.
\end{equation}

\section{Numerical solutions}\label{section 3}
As mentioned in the previous section, the field equations for ${h,f,a}$ \eqref{eq2006} cannot be solved analytically. Therefore, we employ the shooting method to  solve them numerically. We choose to integrate the field equations from horizon to infinity and use the Taylor expansions near the horizon as  initial conditions. There are four parameters ${h_1,f_1,a_1,a_0}$, we set
\begin{equation}
	\label{eq3001}
	a_{0}=0,\quad h_{1}=\frac{1}{r_{0}},\quad f_{1}=\frac{1}{r_{0}}+\delta\,.
\end{equation}
With this parameter setting, we start from a fixed value of the horizon radius $r_0$, and then adjust $\delta$  to satisfy the boundary conditions at infinity. This procedure yields a black hole solution to the equations ~\eqref{eq2006}.

\begin{figure}[H]
	\centering
	\includegraphics[width=0.45\textwidth]{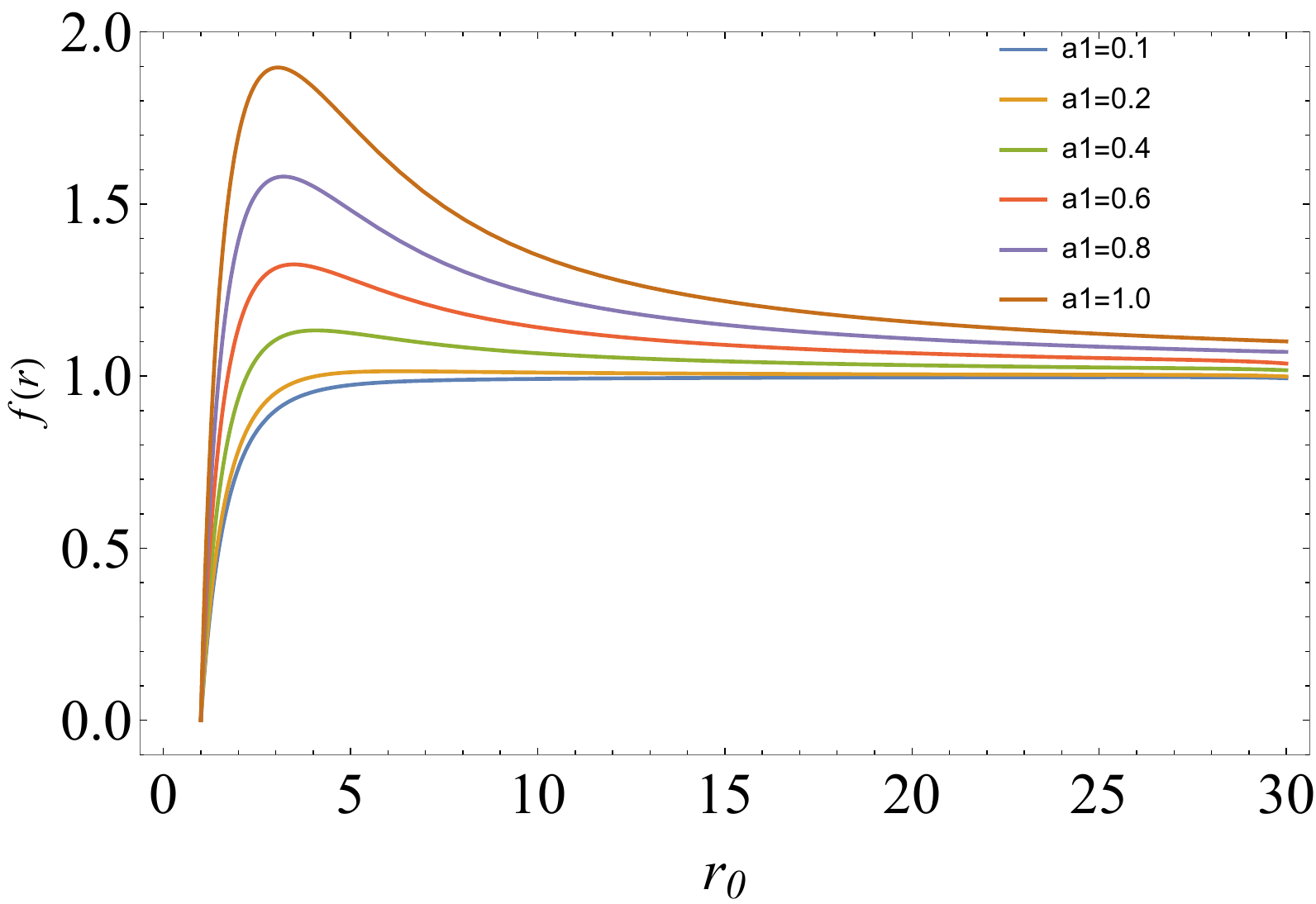}
	\qquad
	\includegraphics[width=0.45\textwidth]{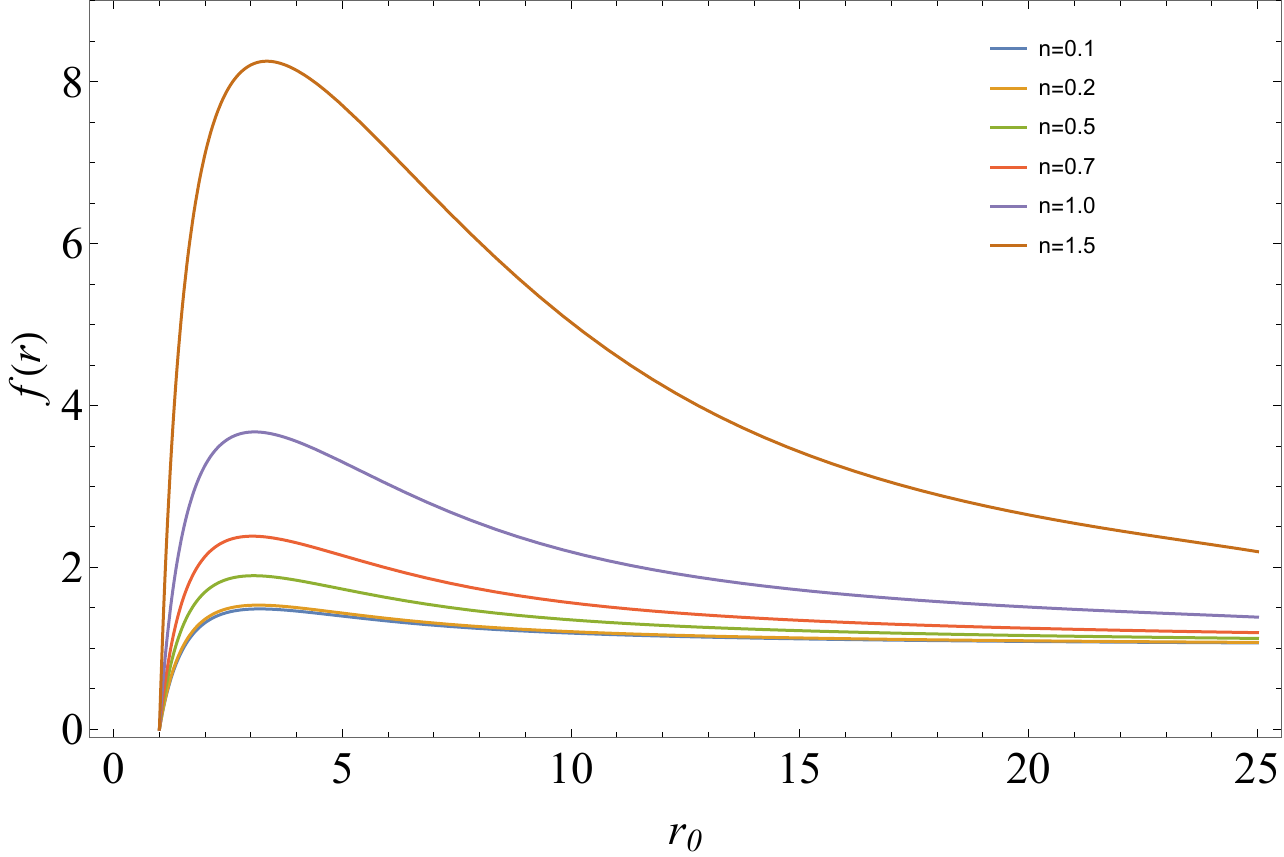}
	\caption{The left panel shows the metric function $f(r)$ for fixed horizon radius $r_{0}=1$, NUT parameter $n=0.5$, and varying values of the electric parameter $a_{1}$. The right panel shows the metric function $f(r)$ with fixed horizon radius $r_{0}=1$, electric parameter  $a_{1}=1.0$, and varying values of the NUT parameter $n$.}
	\label{Fig2901f}
\end{figure}

Figure~\ref{Fig2901f} shows the metric function $f(r)$ for the new black hole solution with the same horizon radius $r_0 = 1$, but different values of  $a_1$ and $n$. The mass parameter m can be extracted from the asymptotic behavior of the metric functions, e.g. $g_{tt} = -1+2 m/r + \dots$. It is worth pointing out that the mass parameter of the new Taub-NUT solutions can be negative, which is a common phenomena for NUT-like black holes. This point can be intuitively seen from the Figure~\ref{Fig2901f}, the curve with a "bump" corresponds to negative mass parameter, while those without a "bump" are associated with positive mass parameter. 

The entropy of the black hole can be further expressed in terms of $\delta$,
\begin{equation}
	\label{eq201704}
	S=\pi(r_{0}^2+n^2)-4\pi \alpha r_{0}\delta.
\end{equation}
It is well-known that the  Gauss-Bonnet combination $\mathcal{G}=R^2-4R_{\mu \nu}R^{\mu \nu}+R_{\mu \nu \rho \sigma}R^{\mu \nu \rho \sigma}$ is a topological term and does not contribute to the equations of motion in four dimensions. However it does contribute a constant to the entropy. We use this freedom to make the entropy proportional to a quarter of horizon radius when $h=f$ or $\delta =0$. 

At the level of solution, the NUT parameter $n$ and electric charge parameter $a_1$ are on equaling footing, spanning a two-dimensional parameter space. In order to fully explore the black hole solutions in this two-dimensional parameter space, we fix one of them (e.g.,$n$) and then vary the other (e.g., $a_1$).

\subsection{Fixed $n$}\label{section 3.1}
First, we fix the NUT parameter to $n=0.5$ and investigate new Taub-NUT black hole solutions for various values of $a_{1}$.

\begin{figure}[H]
	\centering
	\includegraphics[width=0.75\textwidth]{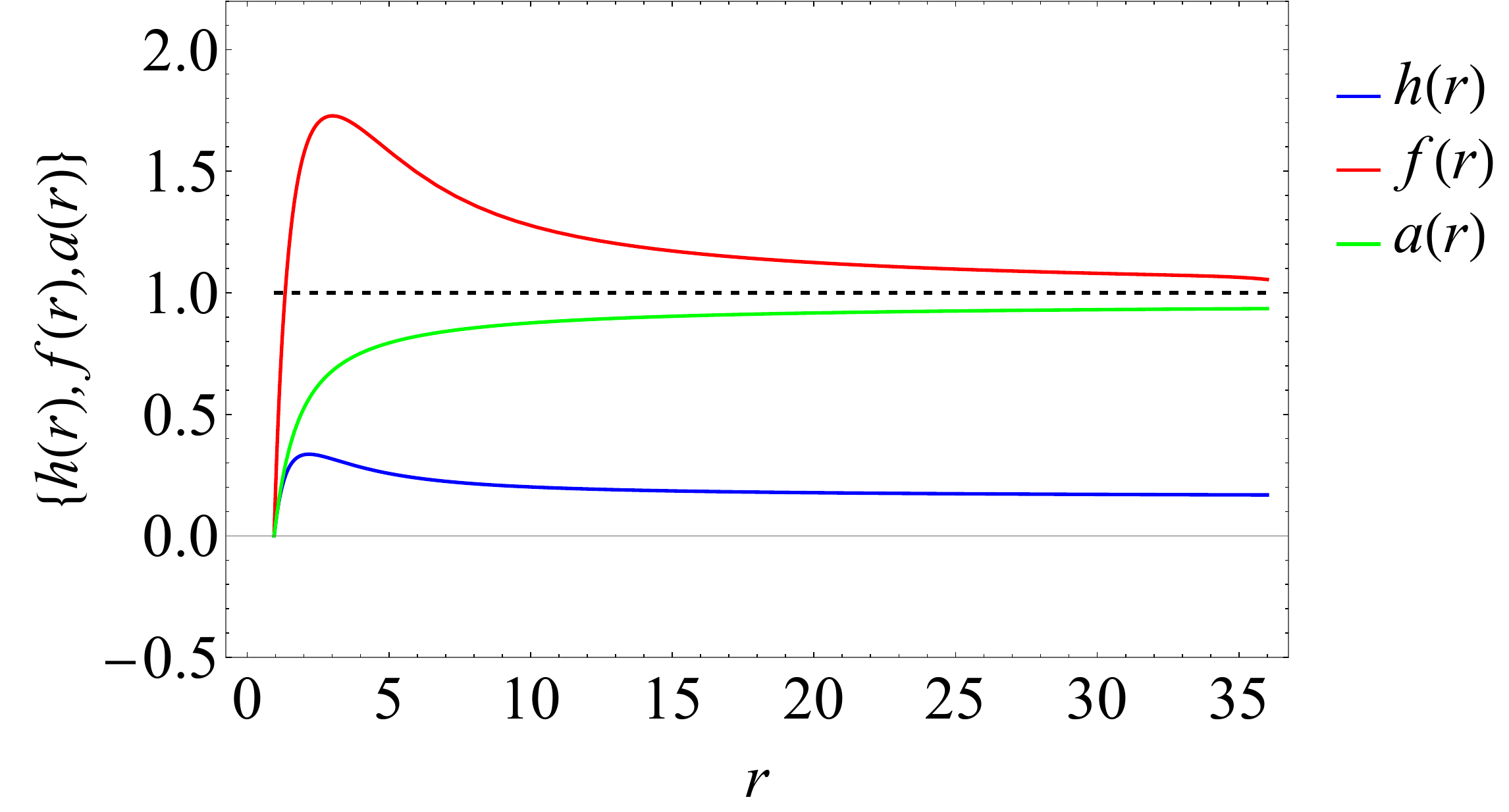}
	\caption{With $a_1=1.0$ and $n=0.5$ fixed, the functions $h(r)$, $f(r)$ and $a(r)$ are shown in the plots for $r_{0}=0.95$. In this case, the integration parameter $\delta$ is positive.}
	\label{Fig3001hfa}
\end{figure}

\begin{figure}[H]
	\centering
	\includegraphics[width=0.45\textwidth]{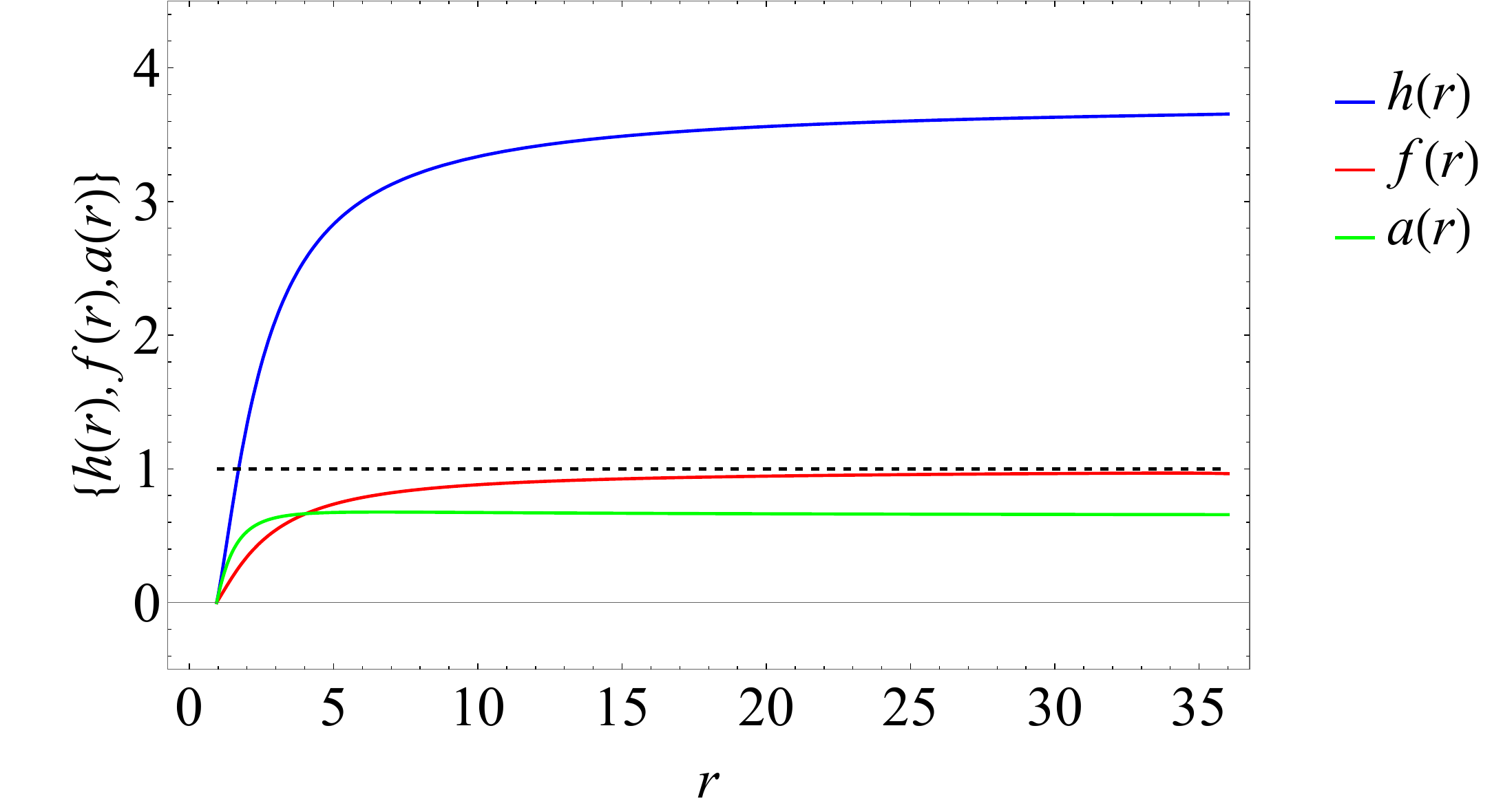}
	\qquad
	\includegraphics[width=0.45\textwidth]{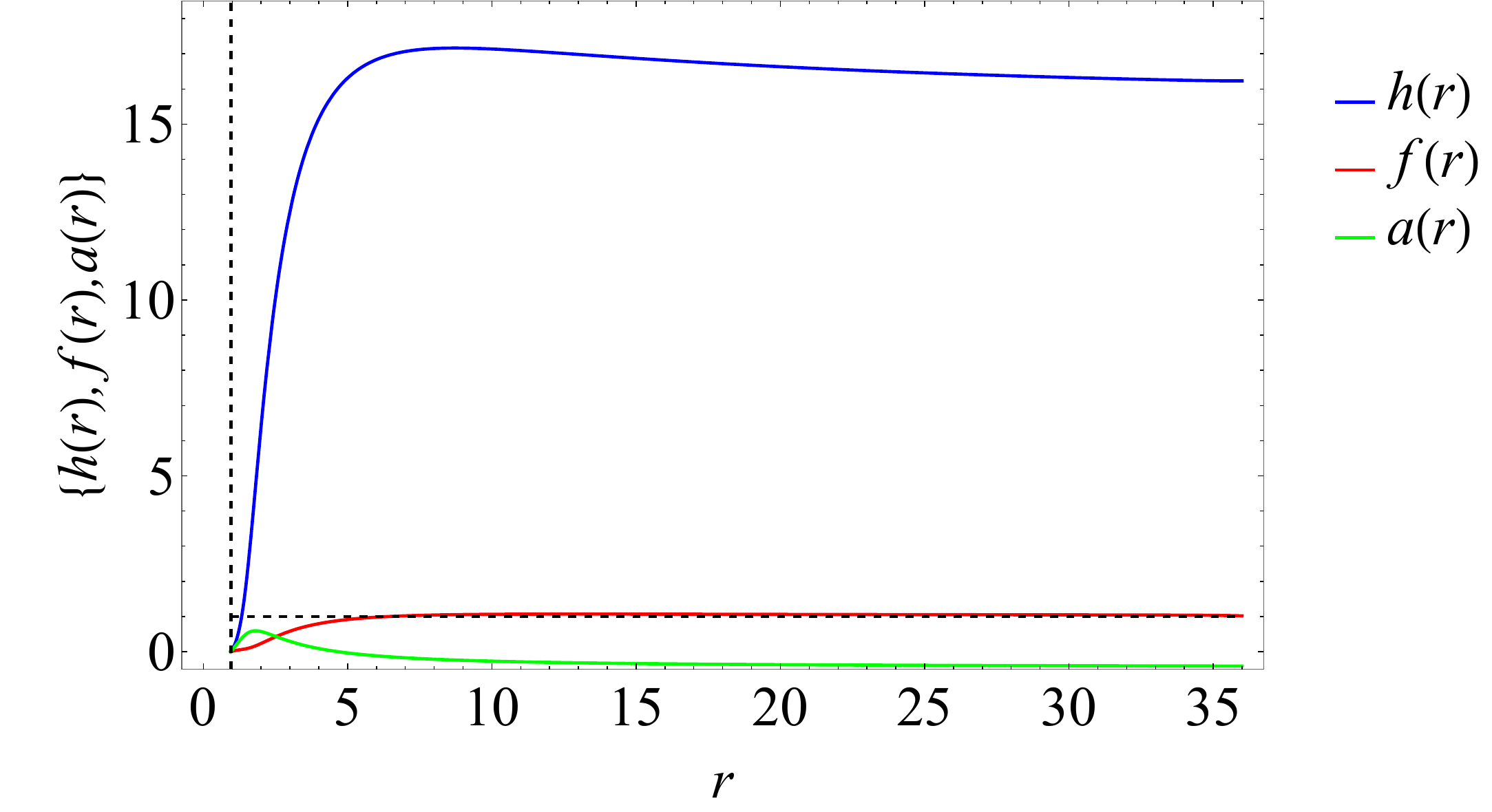}
	\caption{  With $a_{1}=1.0$ and $n=0.5$ fixed, the functions $h(r)$, $f(r)$ and $a(r)$ are shown in the plots for $r_{0}=0.95$. In these two cases, the integration parameter $\delta$ is negative.}
	\label{Fig3001hfa1}
\end{figure}

Setting $a_1=1.0$, and starting from $r_0 =0.95$, we use the shooting method to determine the appropriate value of $\delta$ that satisfies the boundary conditions at infinity. We find that there exist three such values of $\delta$, corresponding to three distinct black hole solutions.
The corresponding metric functions $h(r), f(r)$, and electric potential $a(r)$ are shown in Figure~\ref{Fig3001hfa} and Figure~\ref{Fig3001hfa1}.

For the solution in Figure~\ref{Fig3001hfa}, the integration parameter $\delta$ is positive, and the metric function $f(r)$ exhibits a "bump", indicating that the associated mass parameter is negative.
For the two solutions in Figure~\ref{Fig3001hfa1}, $\delta$ is negative, and the metric functions $f(r)$ show no bump, implying that the mass parameters of these solutions are positive.It can be seen from Figure~\ref{Fig3001hfa} and Figure~\ref{Fig3001hfa1} that metric function $f$ approaches 1 and  $h,a$'s approach a constant at large $r$. 

With these new black hole solutions, we now turn to exploring the thermodynamic properties of these black holes. Two  key thermodynamic quantities, temperature and entropy, whose expressions were derived in the previous section, will be analyzed in detail.

\begin{figure}[H]	
	\centering
	\includegraphics[width=0.4\textwidth]{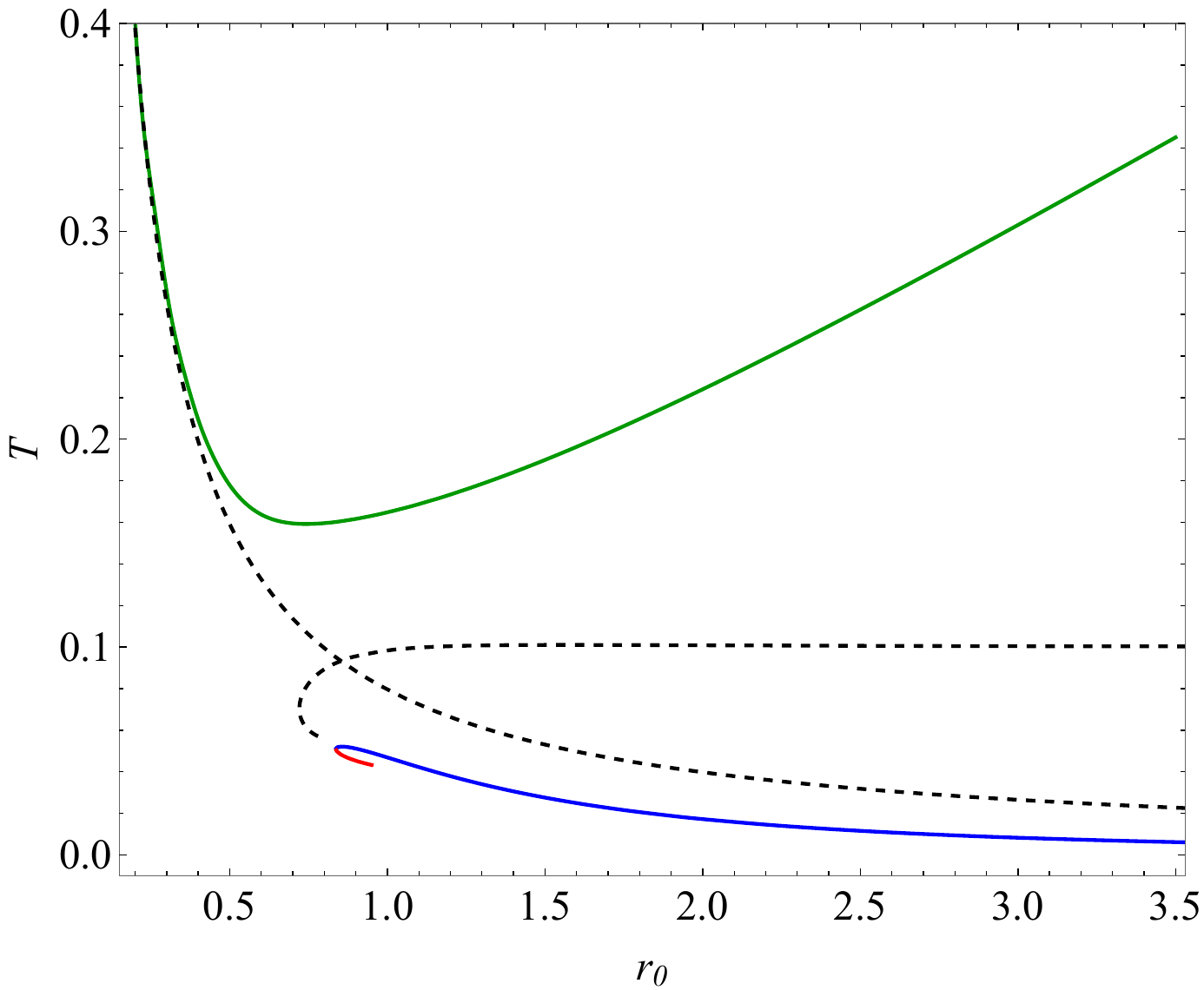}
	\qquad
	\includegraphics[width=0.4\textwidth]{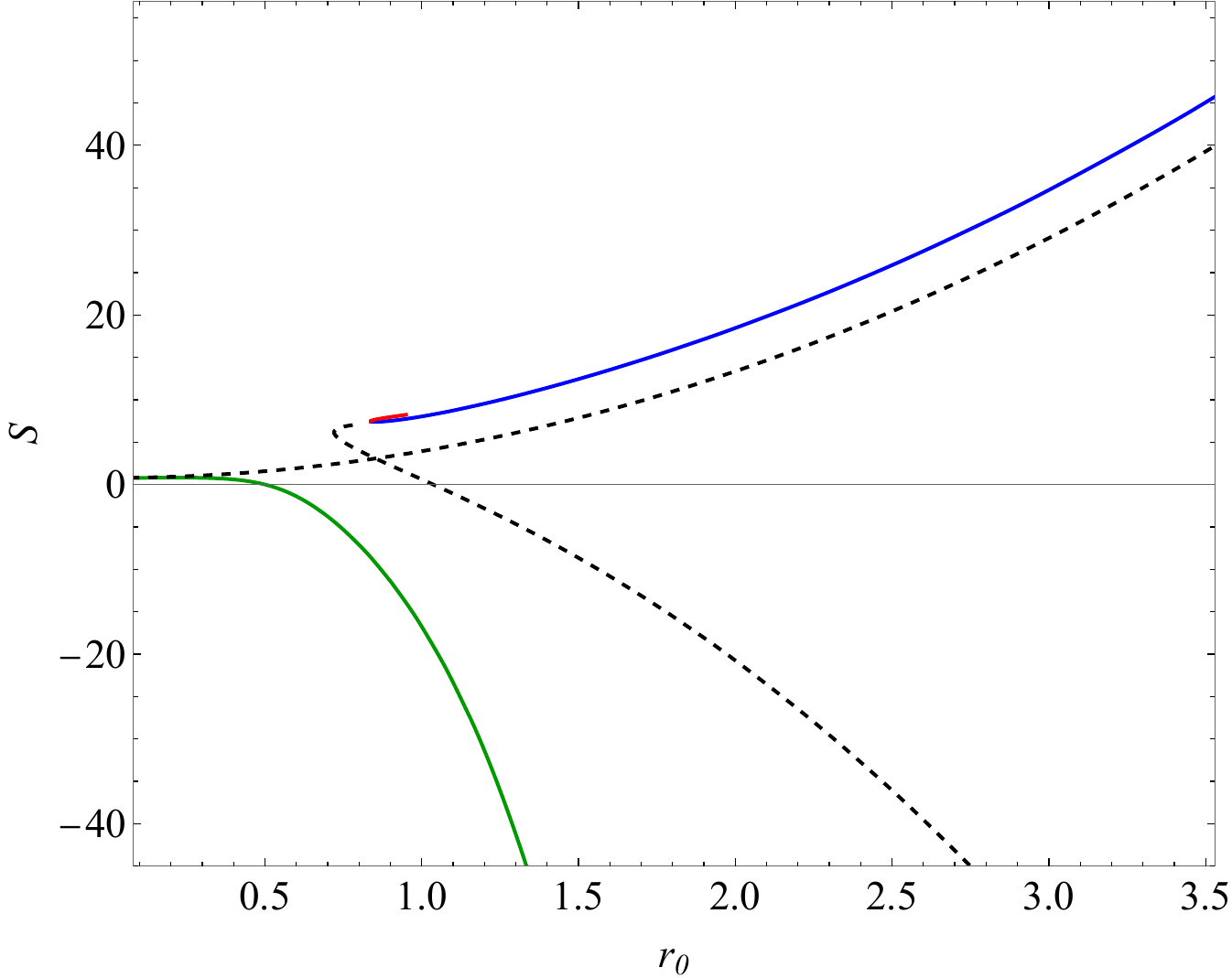}
	\caption{With $a_{1}=1.0$ and $n=0.5$ fixed, the temperature and entropy as functions of $r_{0}$
are shown in the plots. The parameter $\delta$ is positive on the green curves and negative on the blue and red curves. The dashed curves correspond to those of neutral solutions ($a_1=0$).}
	\label{Fig3003}
\end{figure}

As previously noted, there exist three new Taub-NUT-like black hole solutions. Here, we present a comprehensive overview of these three solution branches. In Figure~\ref{Fig3003}, we plot the temperature and entropy as functions of the horizon radius $r_0$. For comparison, we also include the curves corresponding to the neutral Taub-NUT solutions found in~\cite{Chen:2024hsh}.
A significant difference is observed between the charged and neutral solutions: the curves of the neutral solutions intersect, whereas those of the charged solutions remain disconnected.
The parameter $\delta$ is positive for the green solid curve and negative for the red and blue solid curves. Consequently, the temperature of the green curve is higher than that of the other two. In contrast, the entropy of the green curve is lower than that of the red and blue curves.

\begin{figure}[H]	
	\centering
	\includegraphics[width=0.45\textwidth]{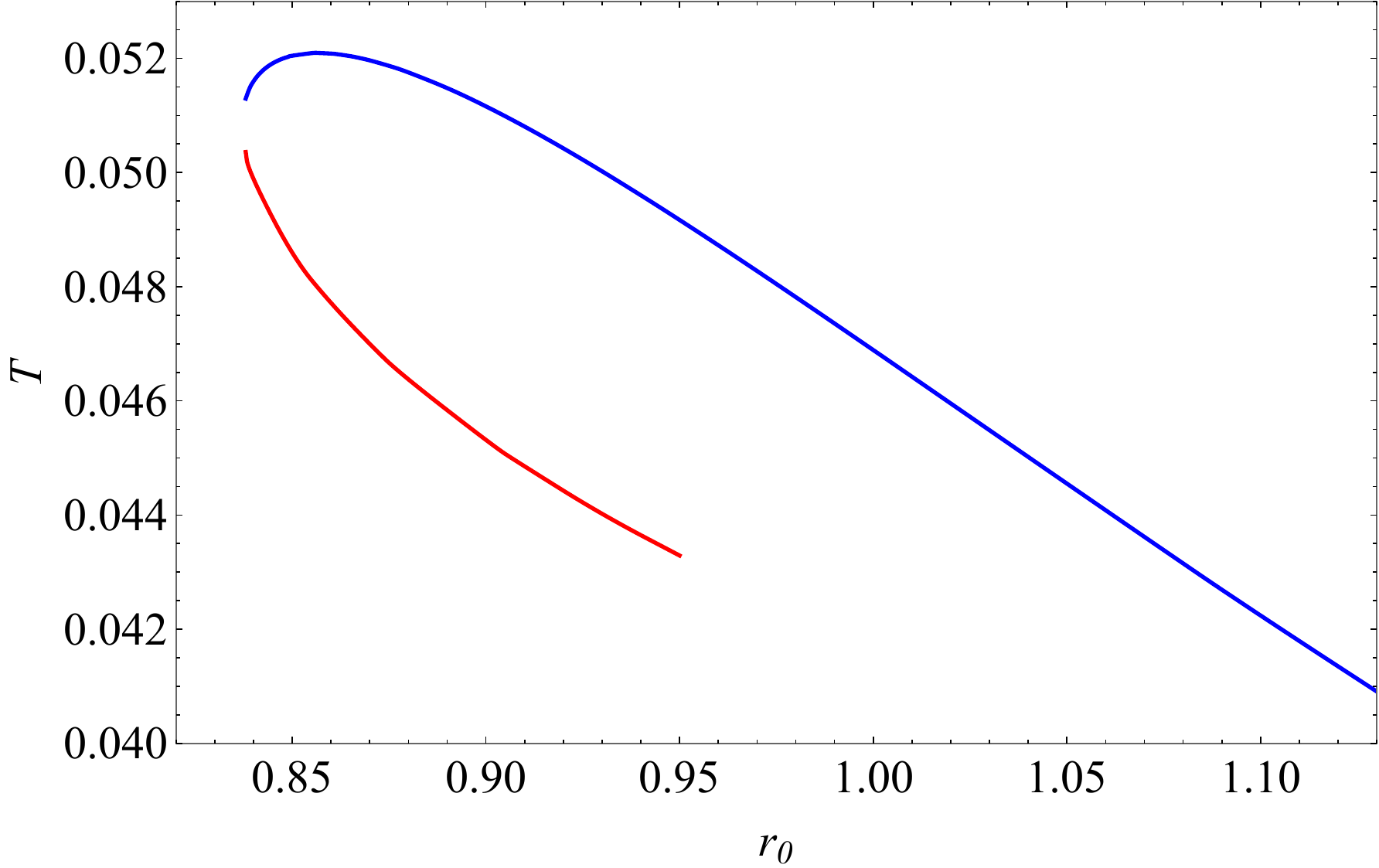}
	\qquad
	\includegraphics[width=0.4\textwidth]{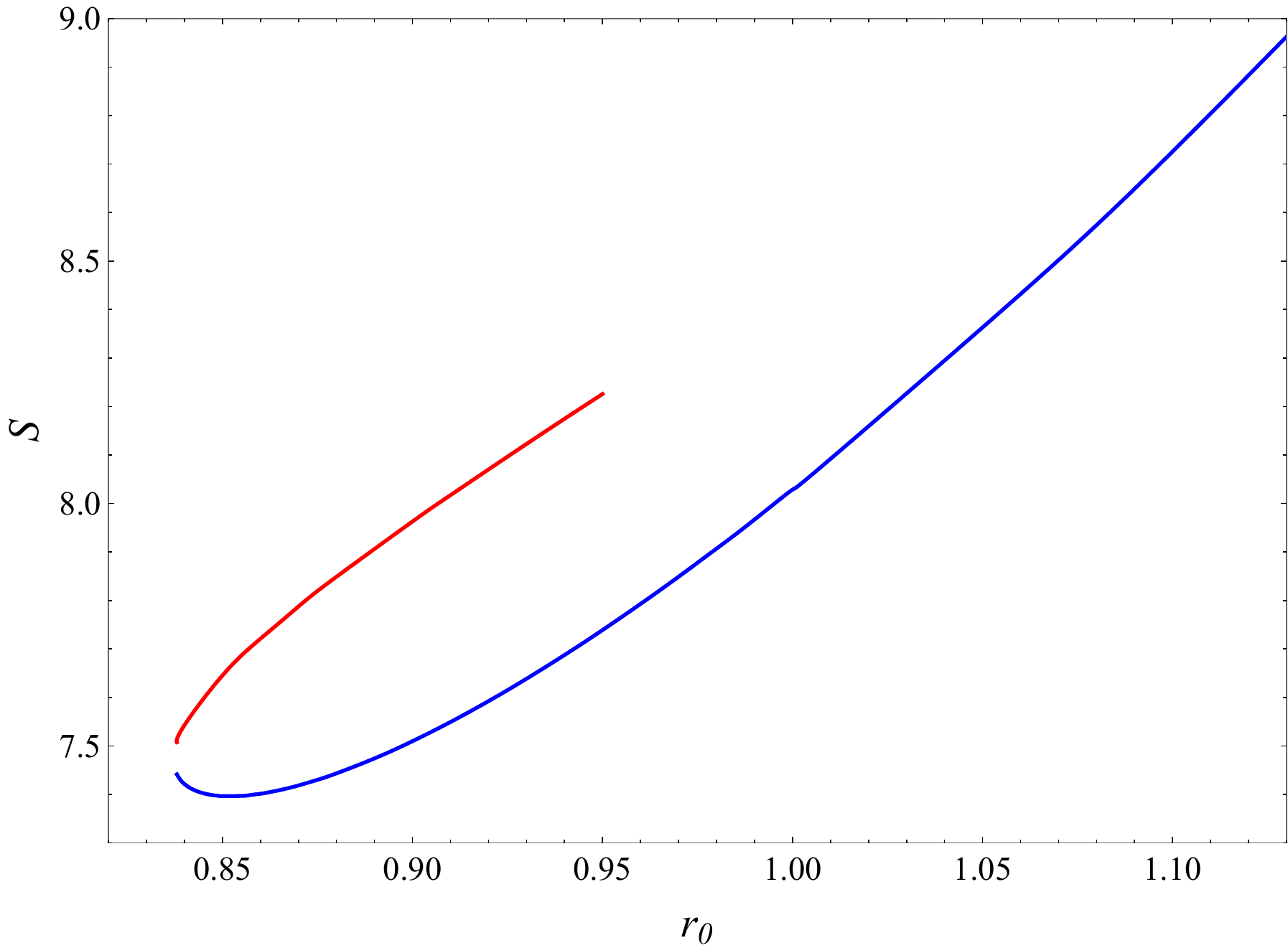}
	\caption{	With $a_{1}=1.0$ and $n=0.5$ fixed, the temperature and entropy as functions of $r_{0}$ are shown in the plots. The parameter $\delta$ is negative on the red and blue curves.}
	\label{Fig30031}
\end{figure}

There exists a minimum horizon radius for the two solutions with negative $\delta$ (the red and blue solid curves), and these two solutions smoothly connect to each other at this minimum radius, as can be seen more clearly in Figure~\ref{Fig30031}. The temperature and entropy curves of these solutions exhibit a "hook" shape at the minimal horizon radius.

\begin{figure}[H]	
	\centering
	\includegraphics[width=0.4\textwidth]{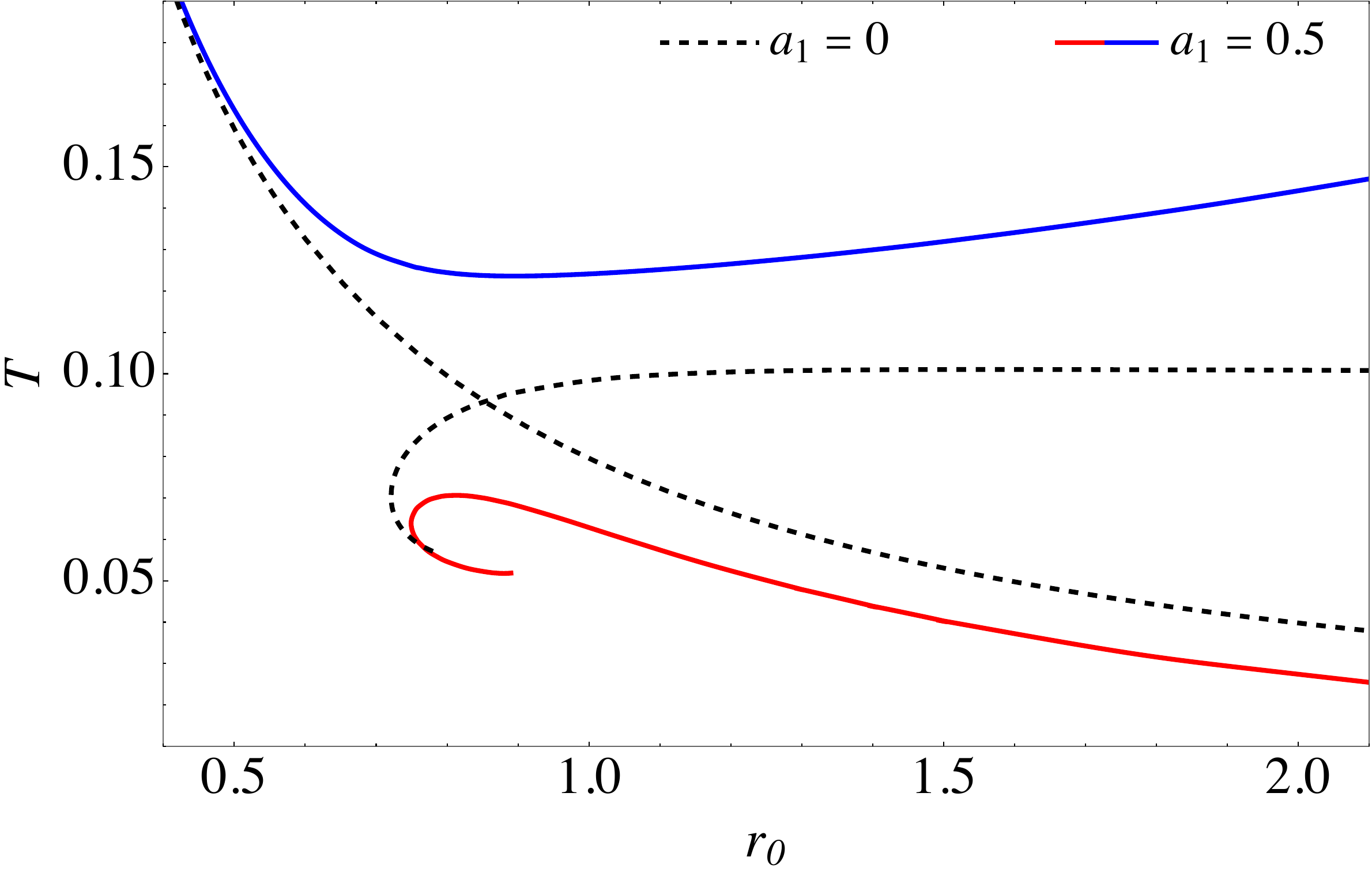}
	\qquad
	\includegraphics[width=0.4\textwidth]{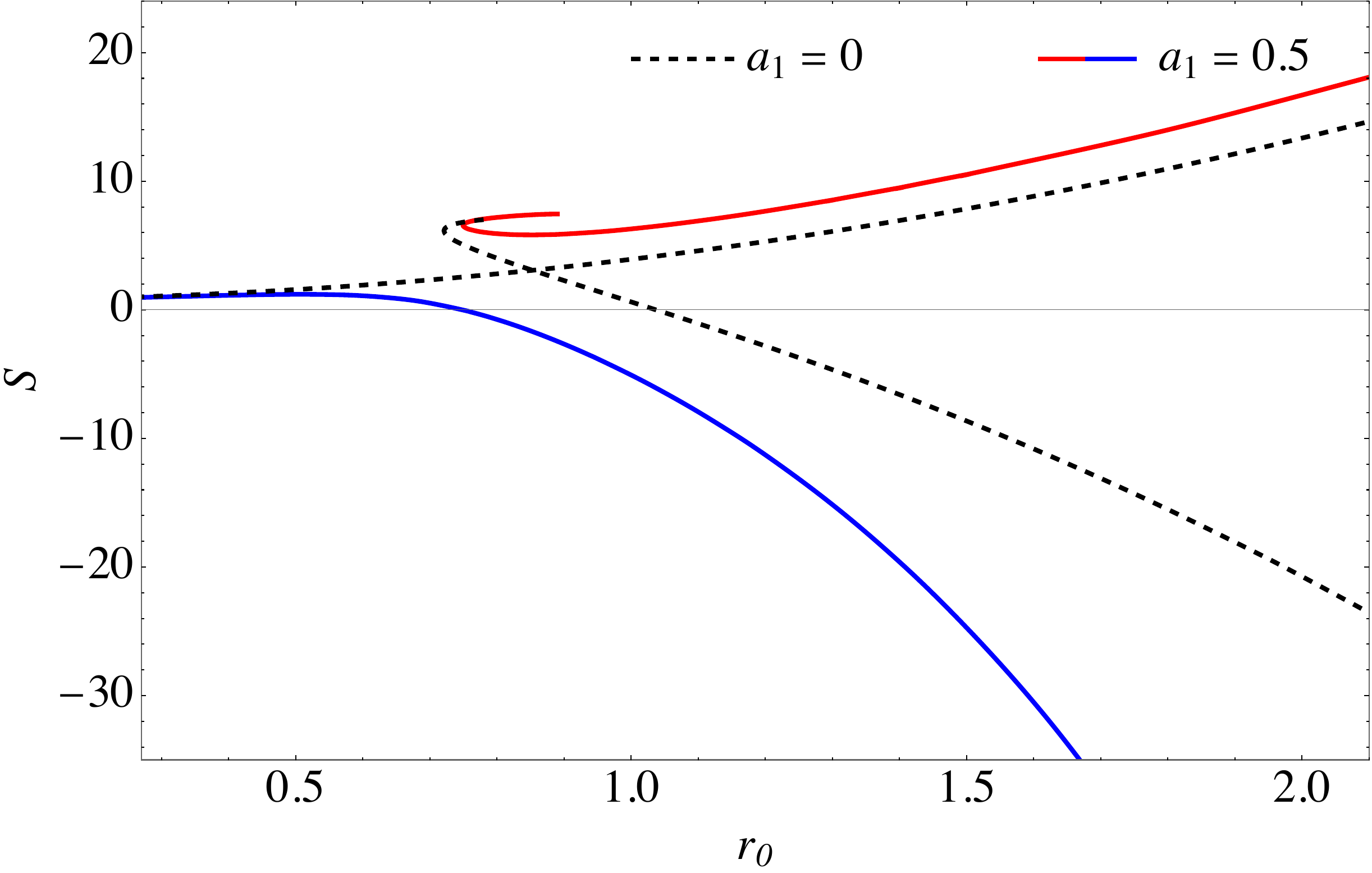}
	\qquad
	\includegraphics[width=0.4\textwidth]{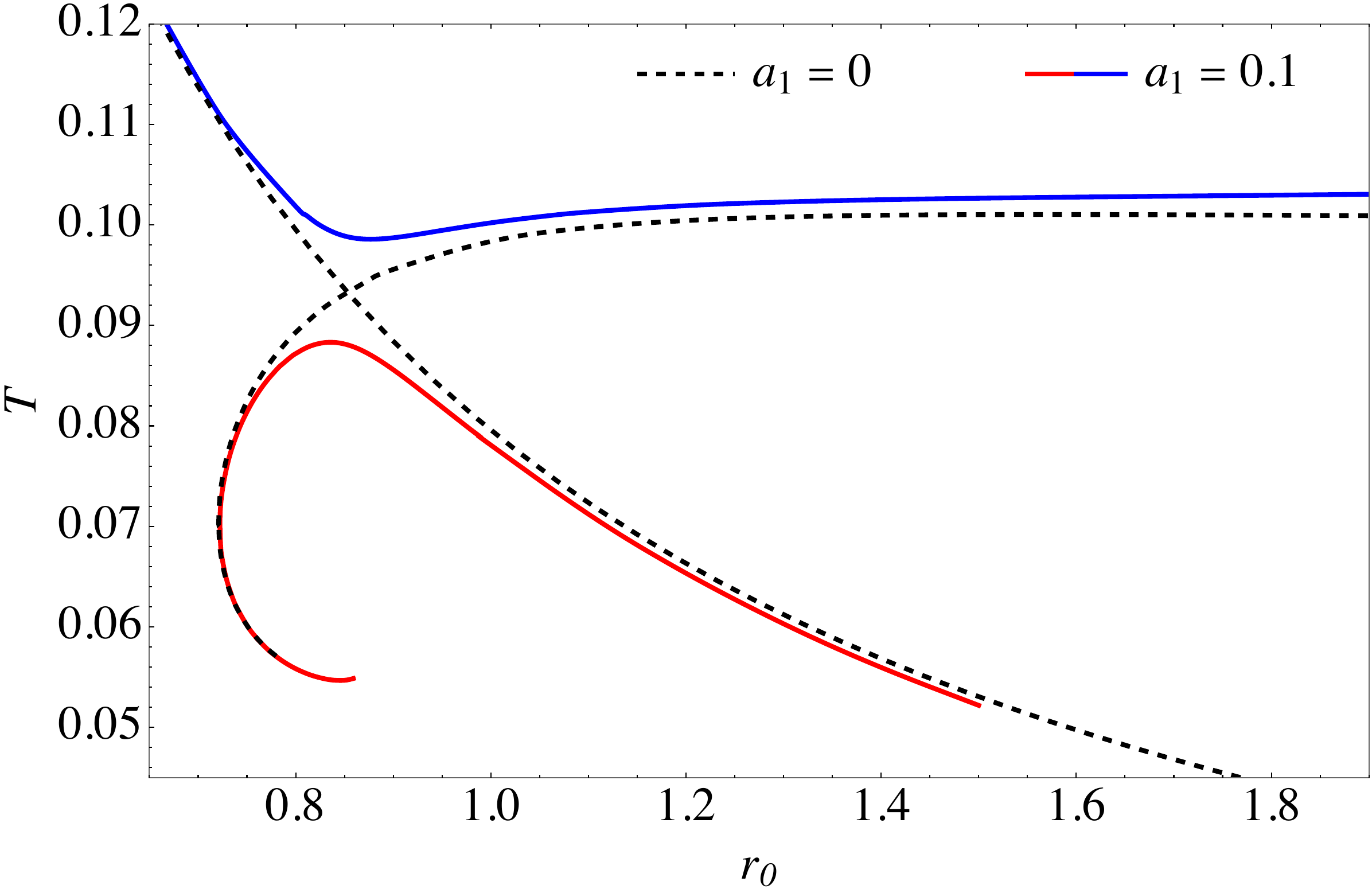}
	\qquad
	\includegraphics[width=0.4\textwidth]{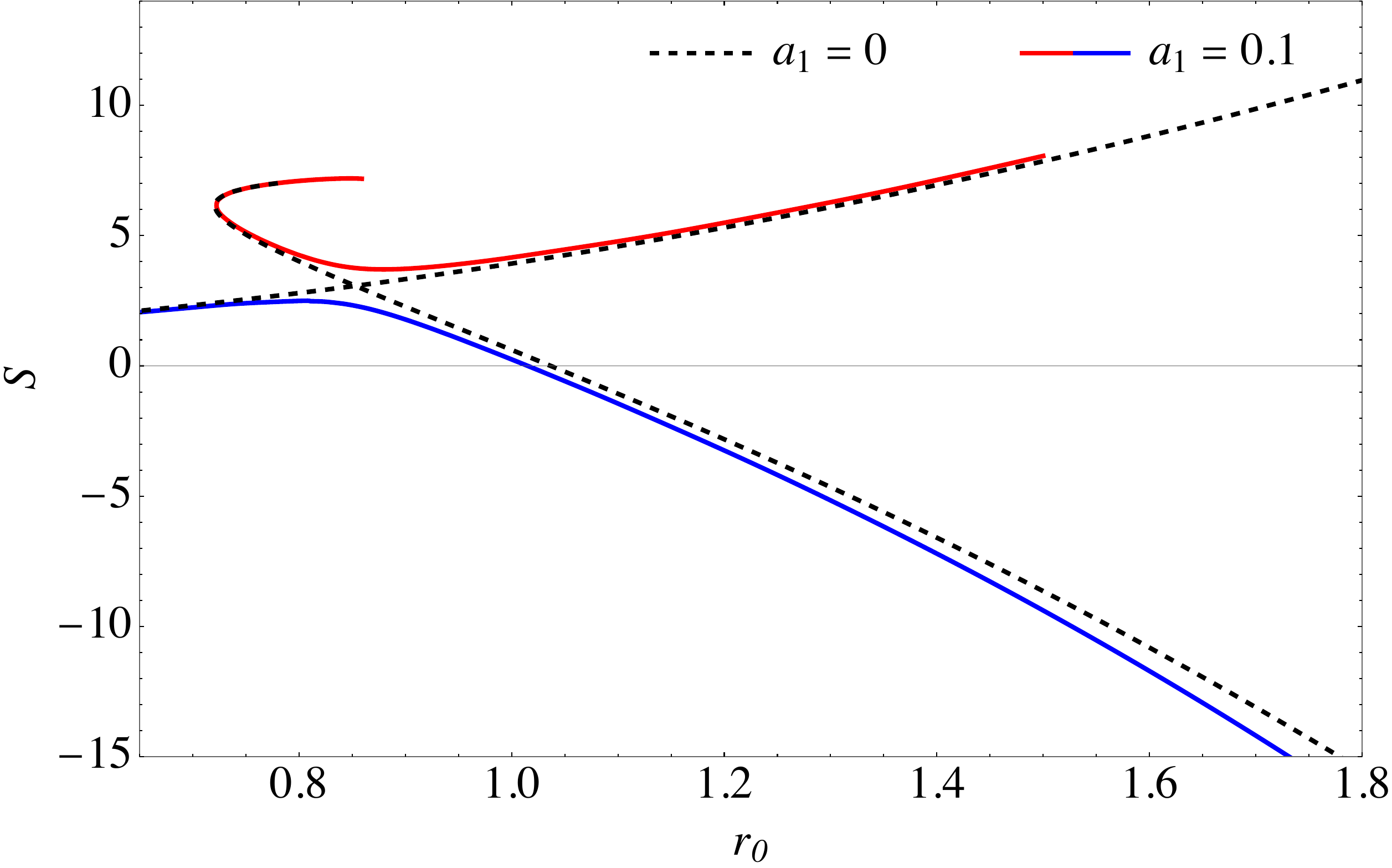}
	\qquad
	\includegraphics[width=0.4\textwidth]{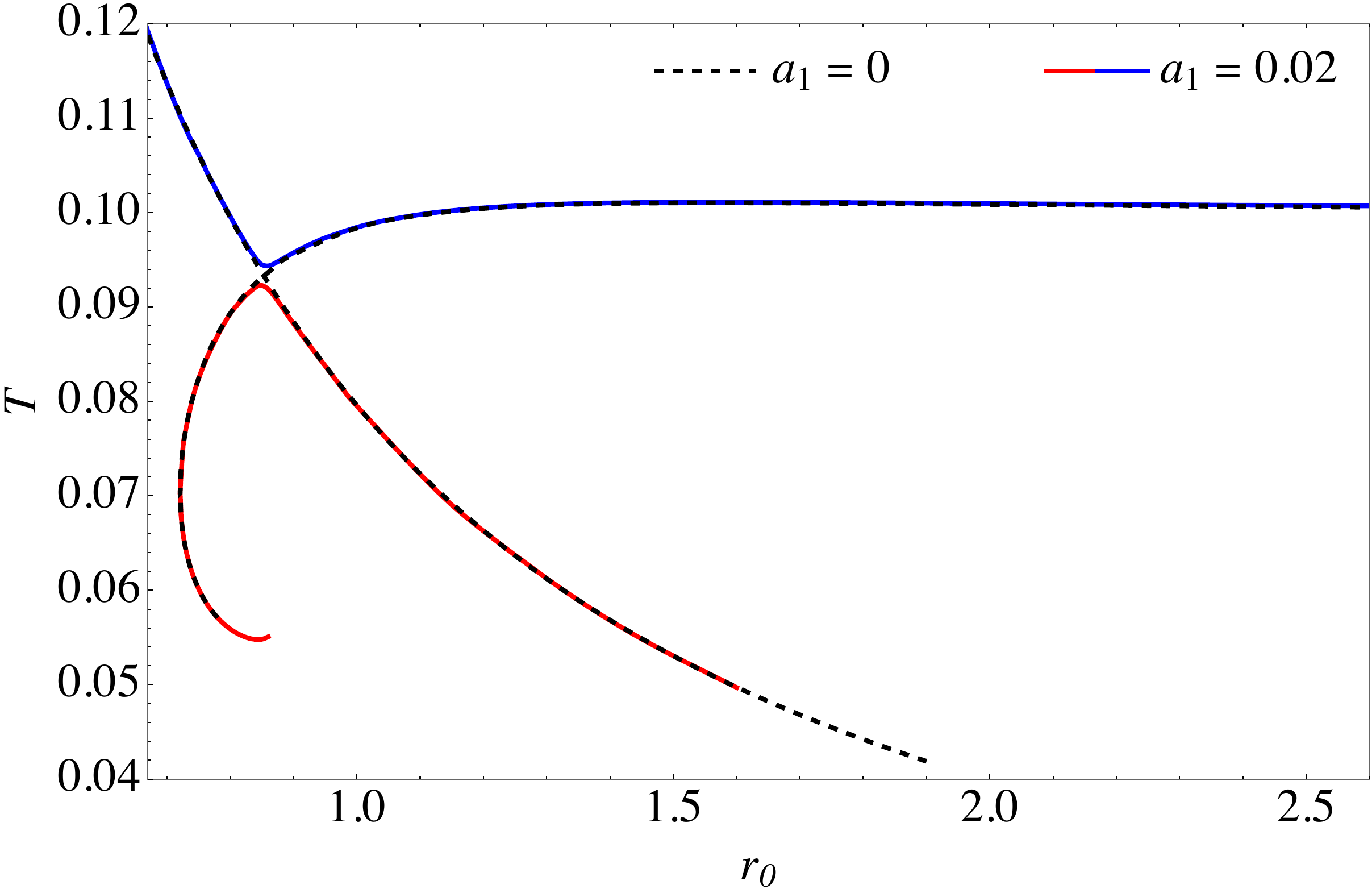}
	\qquad
	\includegraphics[width=0.4\textwidth]{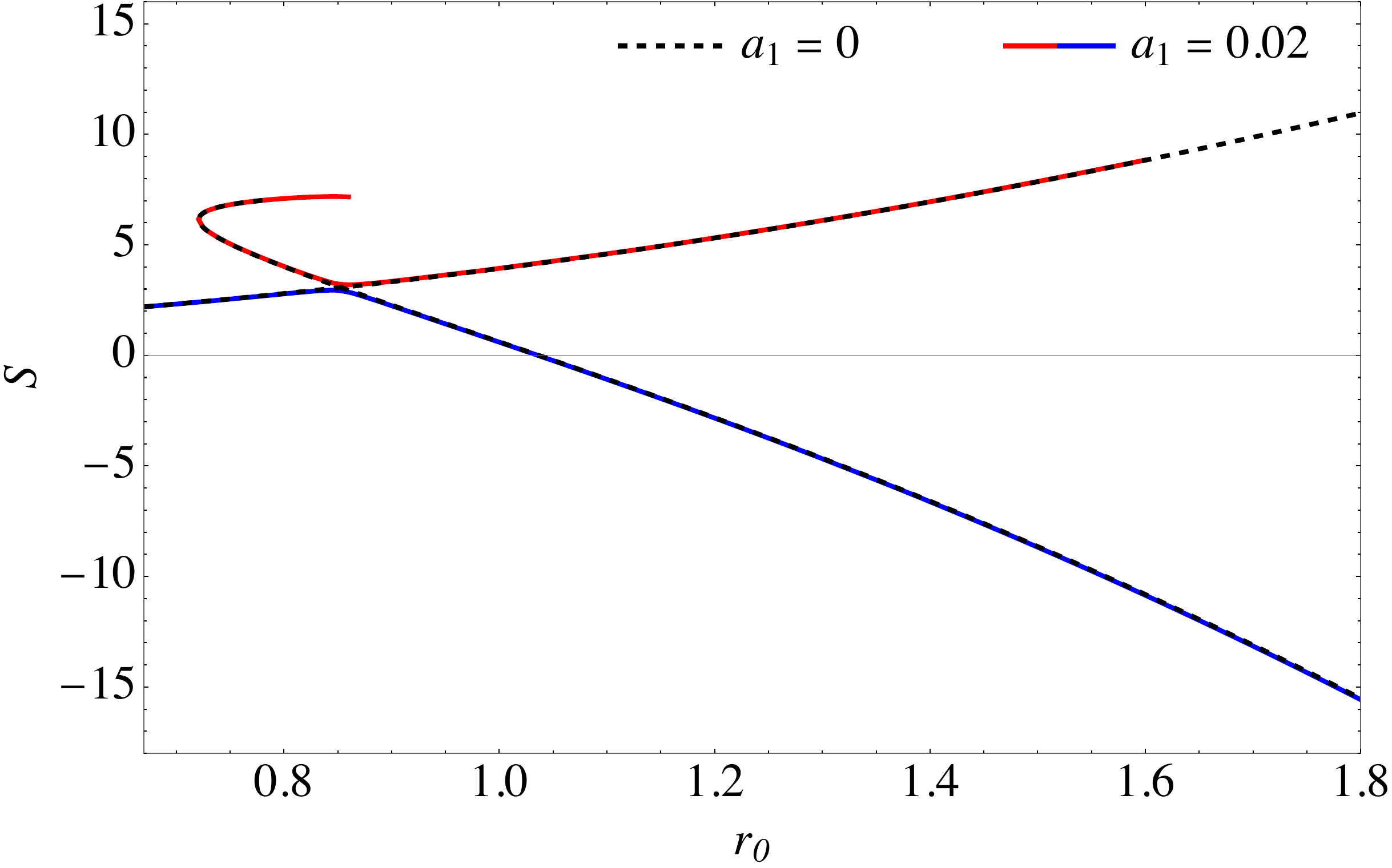}
	\caption{Fixed $n=0.5$ ,$a_{1}=0.5, 0.1, 0.02$. The temperature $T$ and entropy $S$ with different $r_{0}$ are shown in the plots. $\delta$ are positive on the blue curves and negative on the red curves.}
	\label{Fig30035}
\end{figure}

We keep $n=0.5$ and explore new Taub-NUT solutions for various values of~$a_{1} (0.5, 0.1, 0.02)$. The patterns are similar to those for $a_1=1.0$. For each $a_1$, there are three branches of solutions: one with positive $\delta$ and two with negative $\delta$. The two branches with negative $\delta$ share a common minimal horizon radius where they smoothly connect. The temperature $T$ and entropy $S$ as functions of $r_{0}$ are displayed in Figure \ref{Fig30035}. For comparison, we also include the neutral solutions ($a_1=0$) in these figures.

\begin{figure}[H]	
	\centering
	\includegraphics[width=0.85\textwidth]{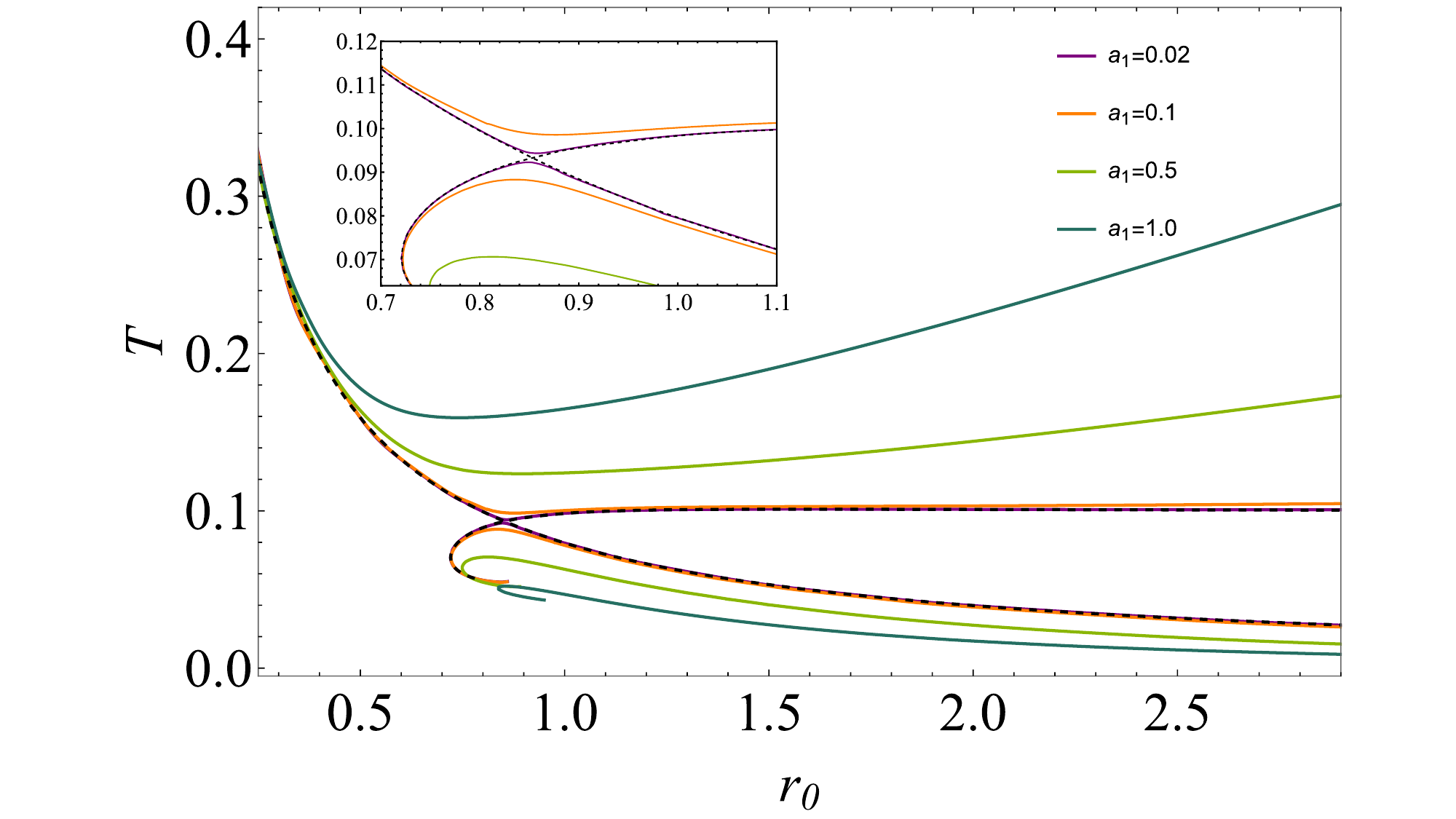}
	\qquad
	\includegraphics[width=0.71\textwidth]{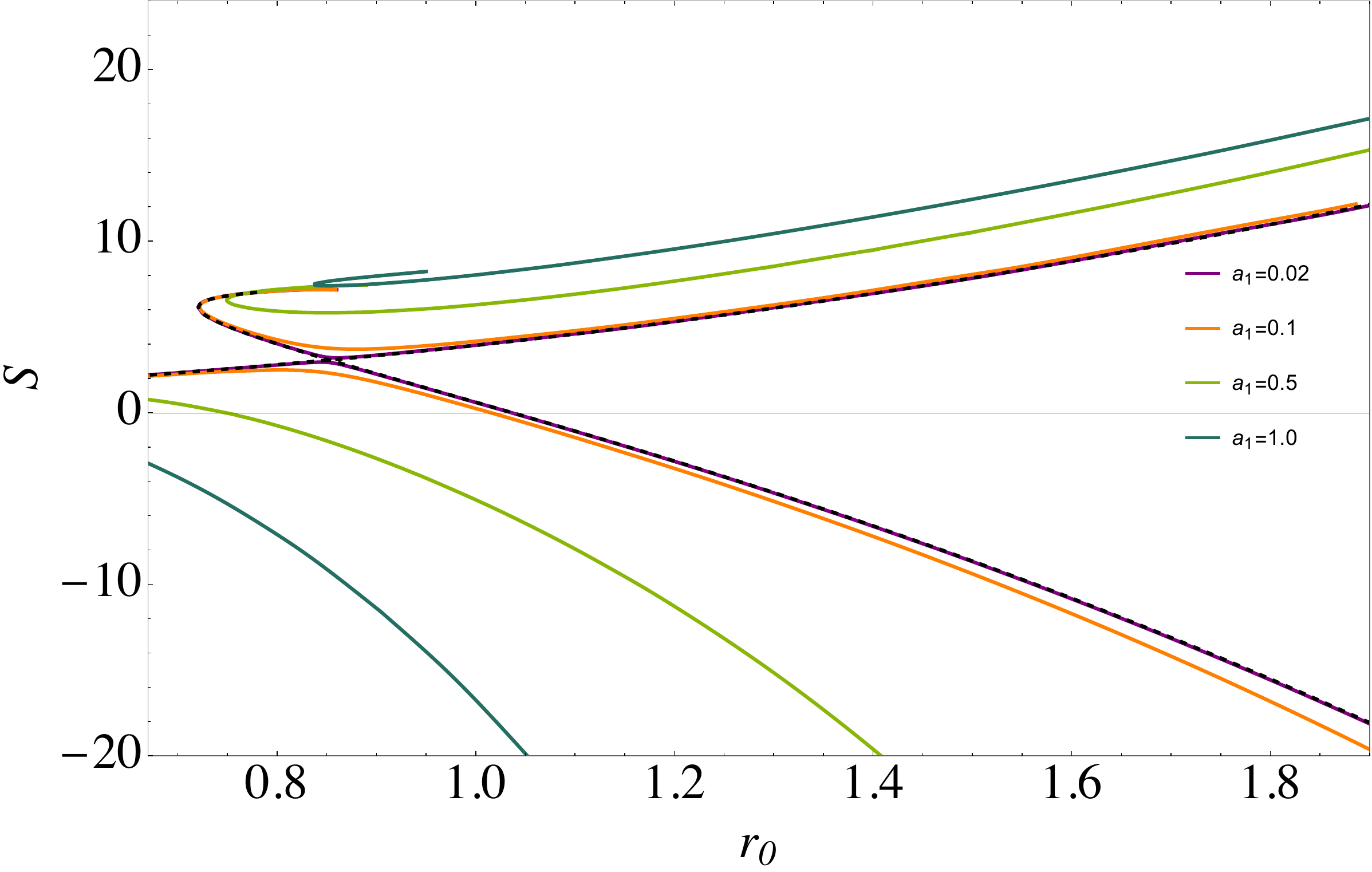}
	\caption{Fixed $n=0.5$, with $a_{1}=0.02,0.1,0.5,1.0$, the temperature and entropy with different $a_{1}$ are shown in the plots. }
	\label{Fig30151}
\end{figure}

From these figures, one can see that the neutral solutions (dashed curves, $a_1=0$) are connected, while our newly found charged solutions are not. As the charge parameter $a_1$
increases, the charged solutions depart further from each other; conversely, when $a_1$
approaches zero, the charged solutions approach the neutral solution. This trend becomes even more apparent when various values of $a_1$ are plotted together, as shown in Figure \ref{Fig30151}.

\subsection{Fixed $a_{1}$}\label{section 3.2}

In this subsection, we fix the parameter $a_{1}=1.0$ and vary the NUT parameter $n$ to study its influence on the black hole solutions.

\begin{figure}[H]	
	\centering
	\includegraphics[width=0.77\textwidth]{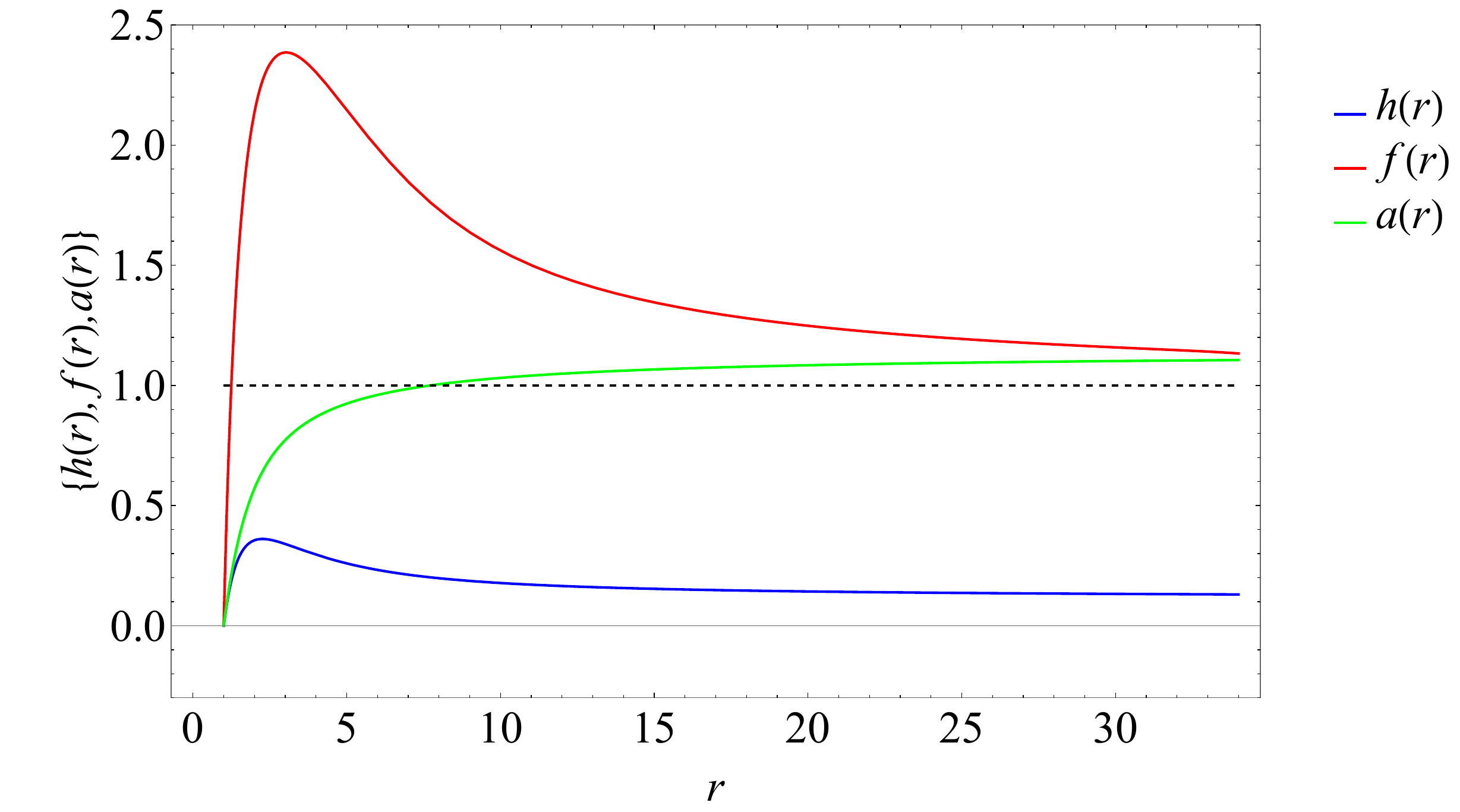}
	\caption{With $a_{1}=1.0$ and $n=0.7$ fixed, the metric functions $h(r)$, $f(r)$, and the electric potential $a(r)$ are shown in the plots for$r_{0}=1.0$. In this case, the integration parameter $\delta$ is positive.
	}
	\label{Fig3016hfa}
\end{figure}

\begin{figure}[H]	
	\centering
	\includegraphics[width=0.47\textwidth]{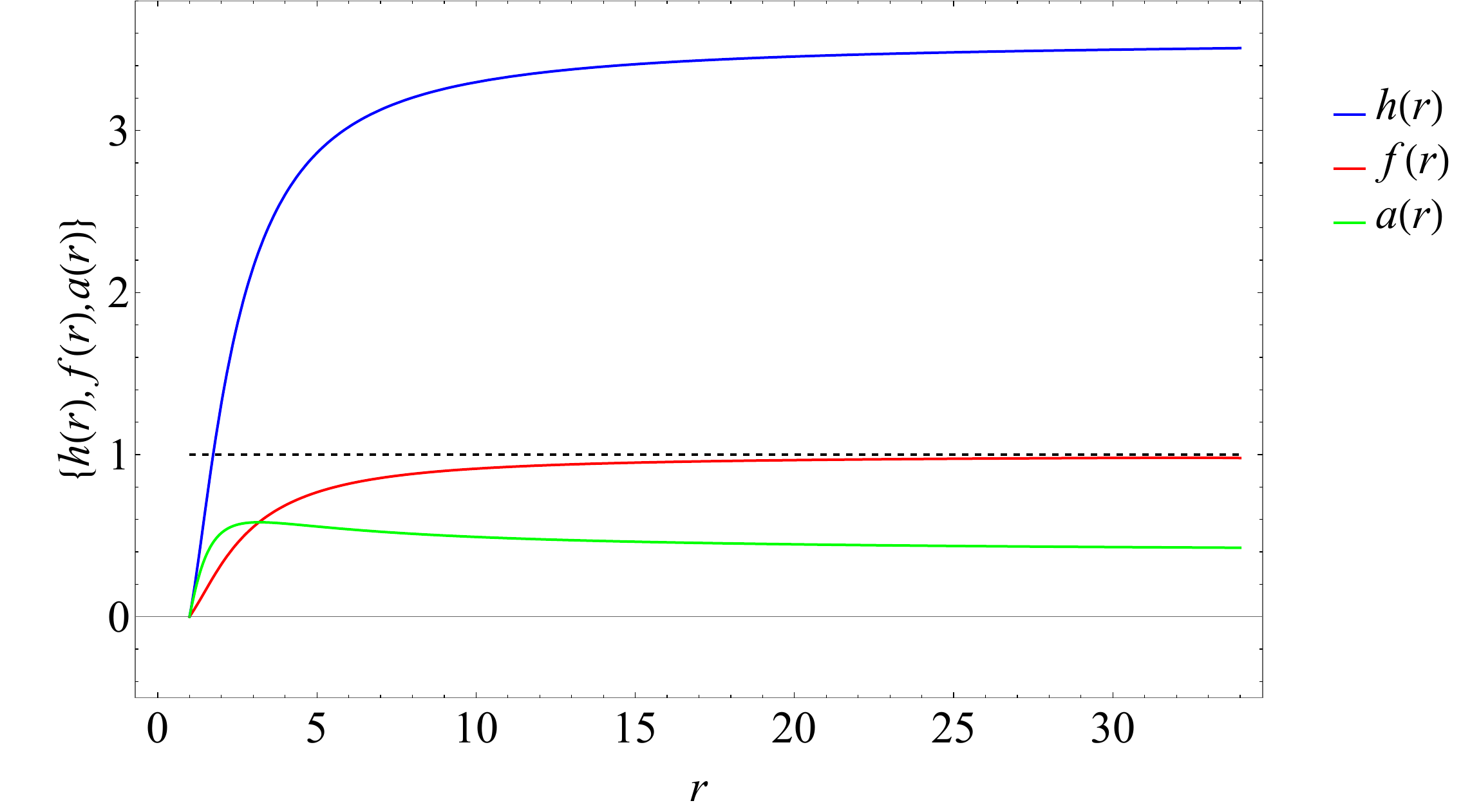}
	\qquad
	\includegraphics[width=0.47\textwidth]{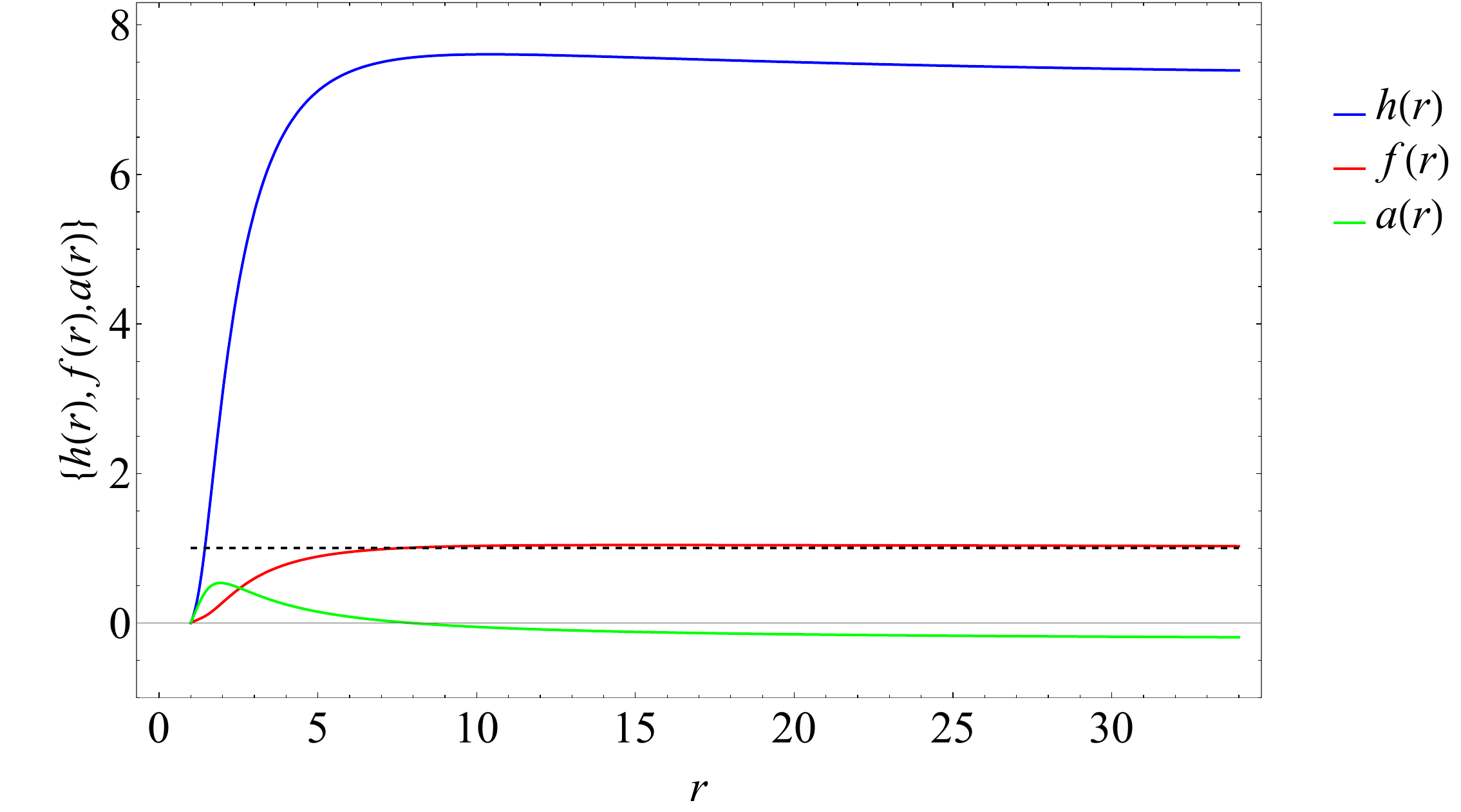}
	\caption{With $a_{1}=1.0$ and $n=0.7$ fixed, the metric functions $h(r)$, $f(r)$, and the electric potential $a(r)$ are shown in the plots for $r_{0}=1.0$. In this case, the integration parameter $\delta$ is negative.}
	\label{Fig3016hfa1}
\end{figure}

Initially, we set $a_{1}=1.0$ and $n=0.7$. We find that the behavior is similar to the case with $a_{1}=1.0$ and $n=0.5$: there are three branches of solutions, one with positive $\delta$ and two with negative $\delta$. Setting $r_{0}=0.95$, the metric functions and electric potential of the three branches are depicted in Figures \ref{Fig3016hfa} and \ref{Fig3016hfa1}. Among these, two solutions correspond to a negative mass parameter, while the remaining one has a positive mass parameter.

\begin{figure}[H]	
 	\centering
 	\includegraphics[width=0.4\textwidth]{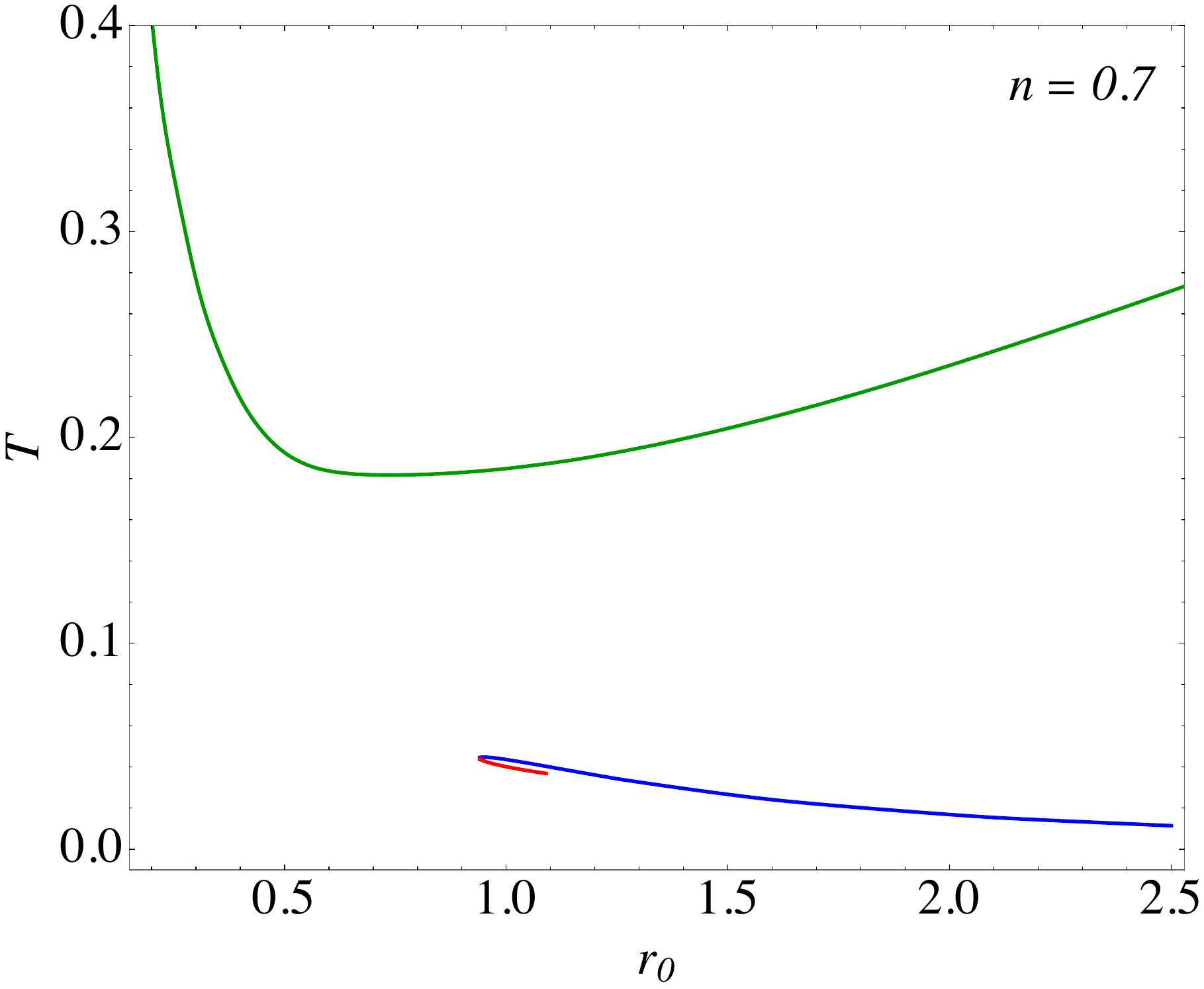}
 	\qquad
 	\includegraphics[width=0.4\textwidth]{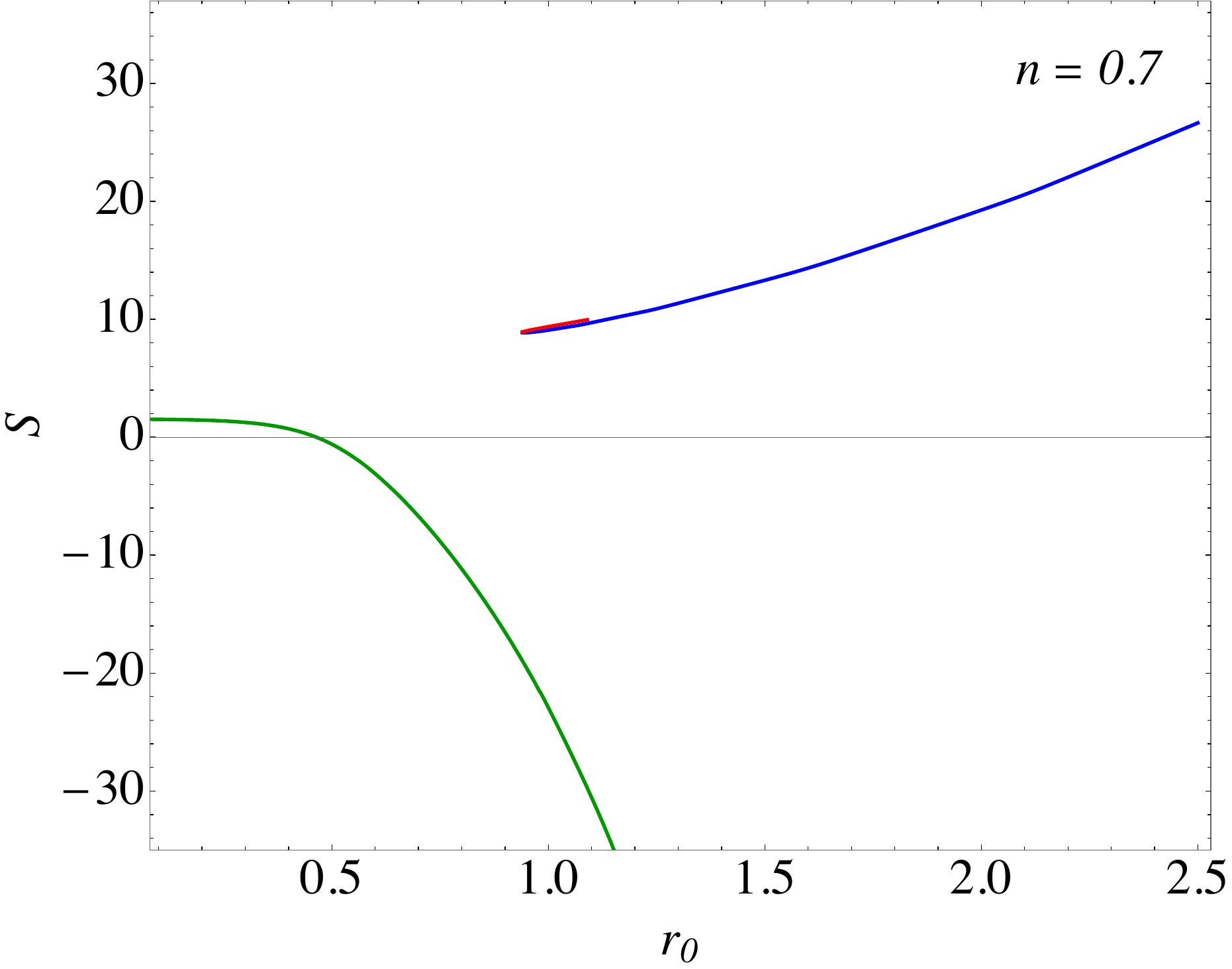}
 	\caption{With $a_{1}=1.0$ and $n=0.7$ fixed, the temperature and entropy as functions of $r_{0}$
 	are shown in the plots.	
 	}
 	\label{Fig3018}
 \end{figure}

The temperature and entropy as functions of the horizon radius $r_0$ for the three branches of solutions are shown in Figure~\ref{Fig3018}. These branches fall into two groups: the one with positive $\delta$ is clearly separated from the other two with negative $\delta$.

\begin{figure}[H]	
	\centering
	\includegraphics[width=0.4\textwidth]{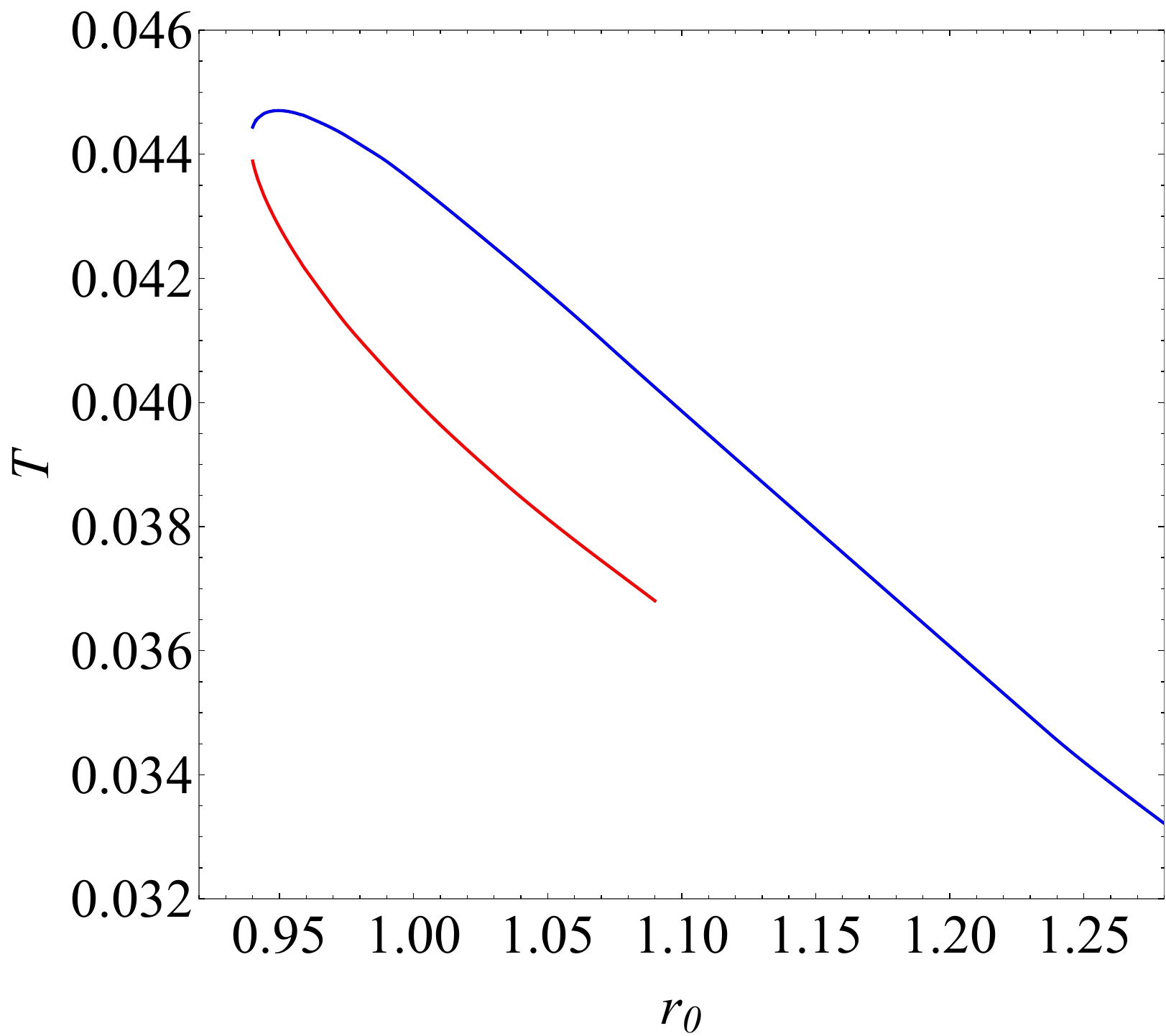}
	\qquad
	\includegraphics[width=0.4\textwidth]{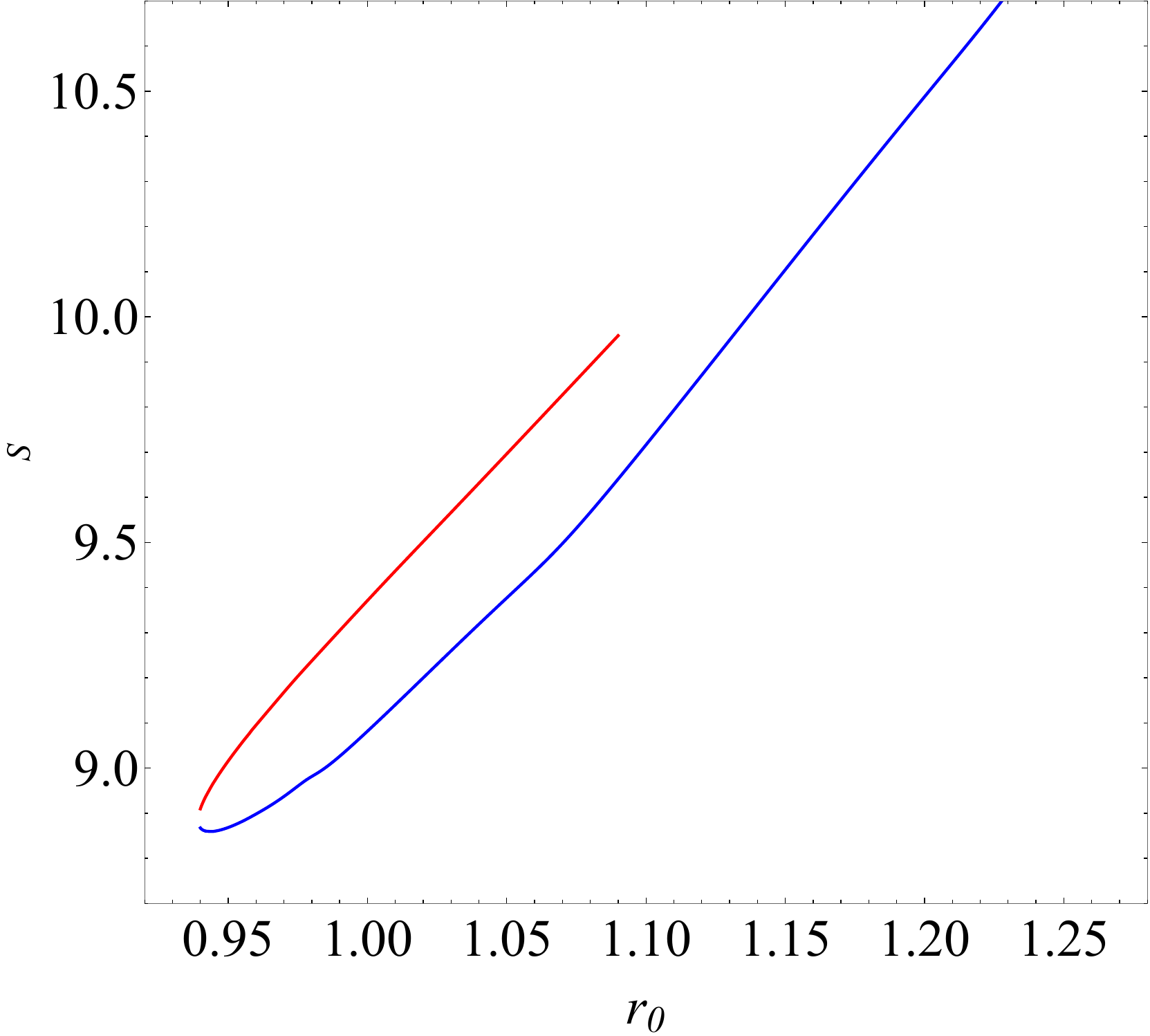}
	\caption{With $a_{1}=1.0$ and $n=0.7$ fixed, the temperature and entropy of the solutions with negative $\delta$ are shown in the plots.	}
	\label{Fig30181}
\end{figure}

 The two solutions with negative $\delta$ are difficult to distinguish in Figure~\ref{Fig3018}. To clarify, we plot the temperature and entropy curves of these two solutions in Figure~\ref{Fig30181}. It is now easy to observe that these two solutions share the same minimal horizon radius, where they smoothly connect to each other. Moreover, the solution with the lower temperature has a shorter extent in the plot. 

\begin{figure}[H]	
	\centering
	\includegraphics[width=0.77\textwidth]{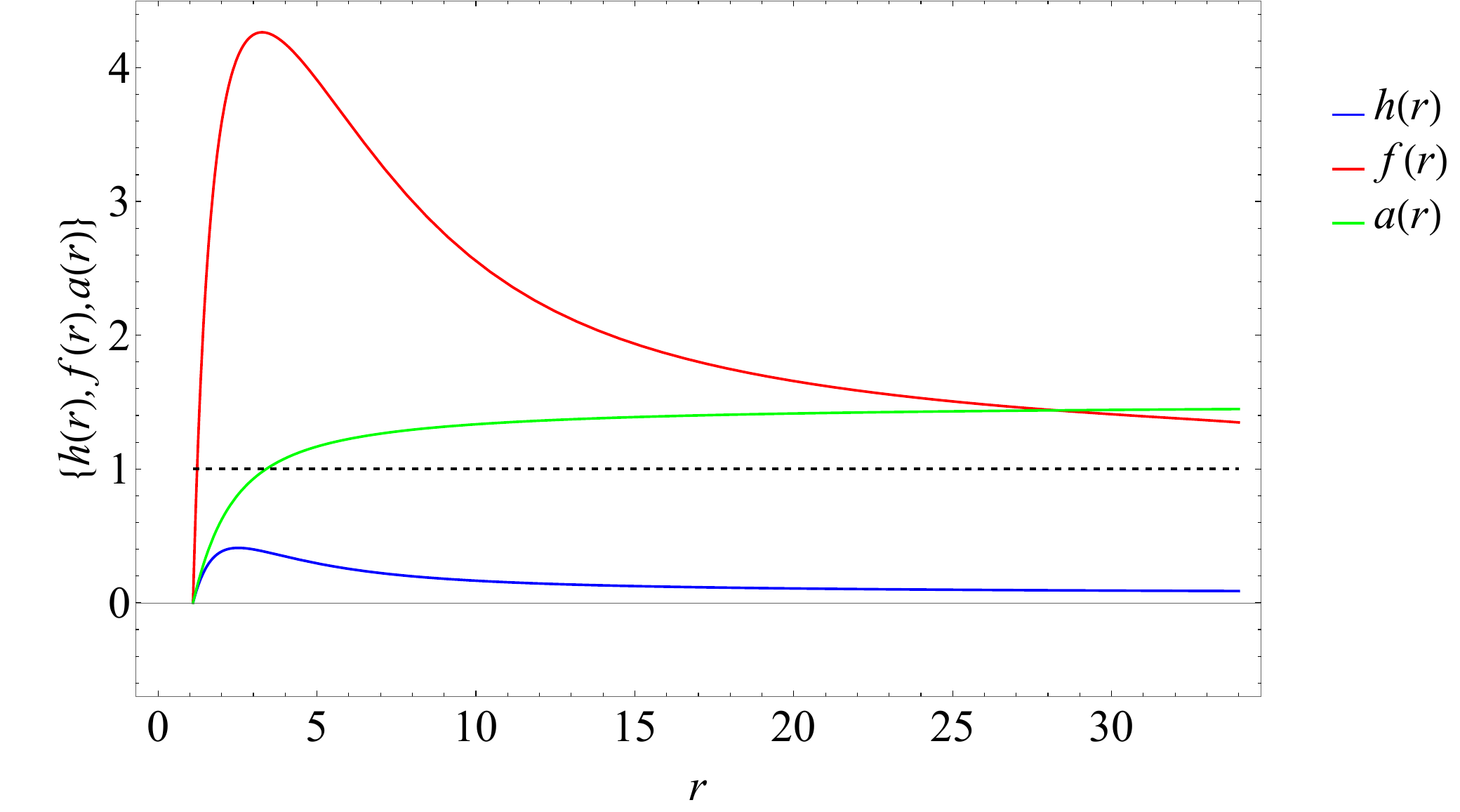}
	\caption{With $a_{1}=1.0$ and $n=1.0$ fixed, the metric functions $h(r), f(r)$, and the electric potential $a(r)$ for the solution with positive $\delta$ are shown in the plot for $r_{0}=1.1$.	}
	\label{Fig3019hfa}
\end{figure}

\begin{figure}[H]	
	\centering
	\includegraphics[width=0.47\textwidth]{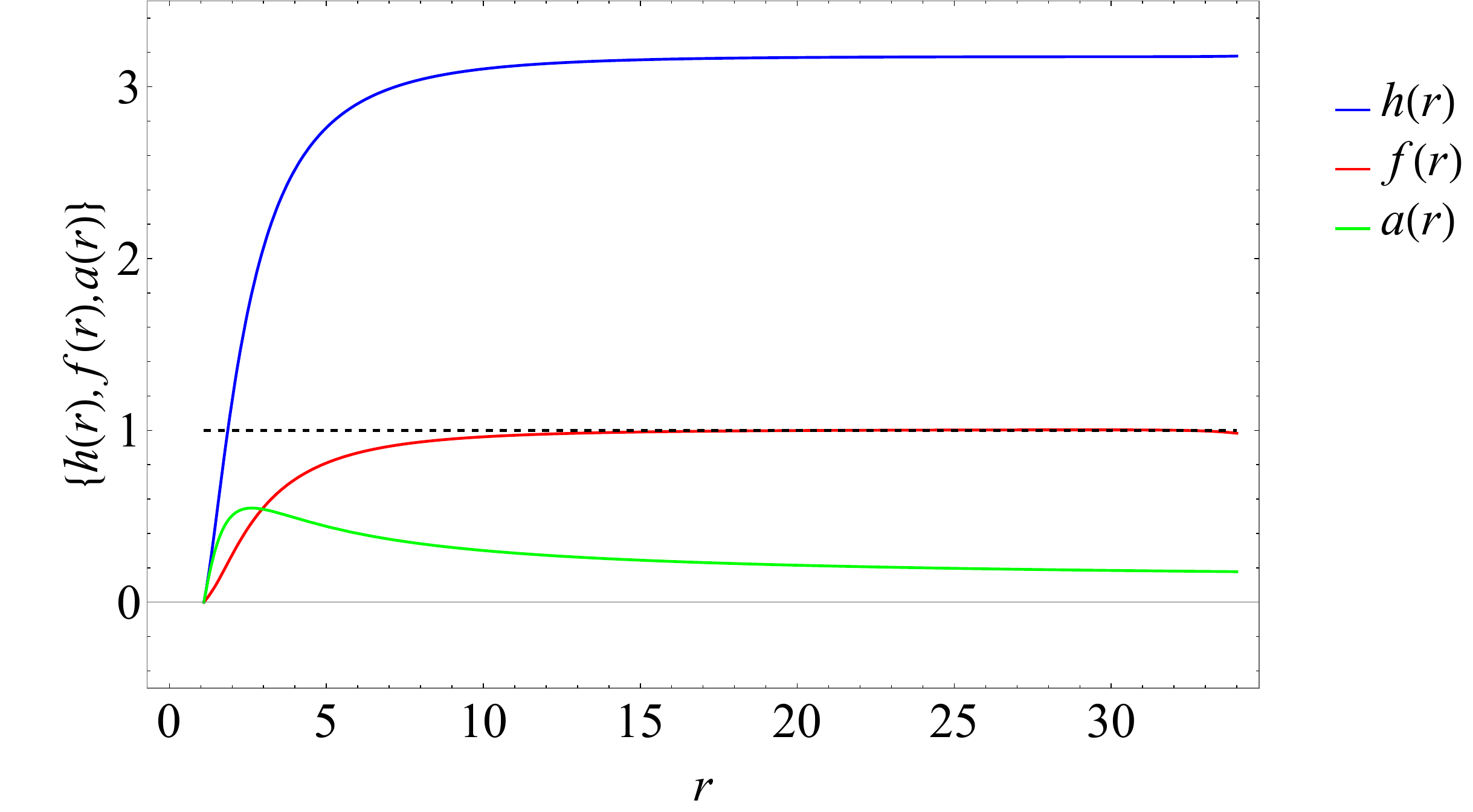}
	\qquad
	\includegraphics[width=0.47\textwidth]{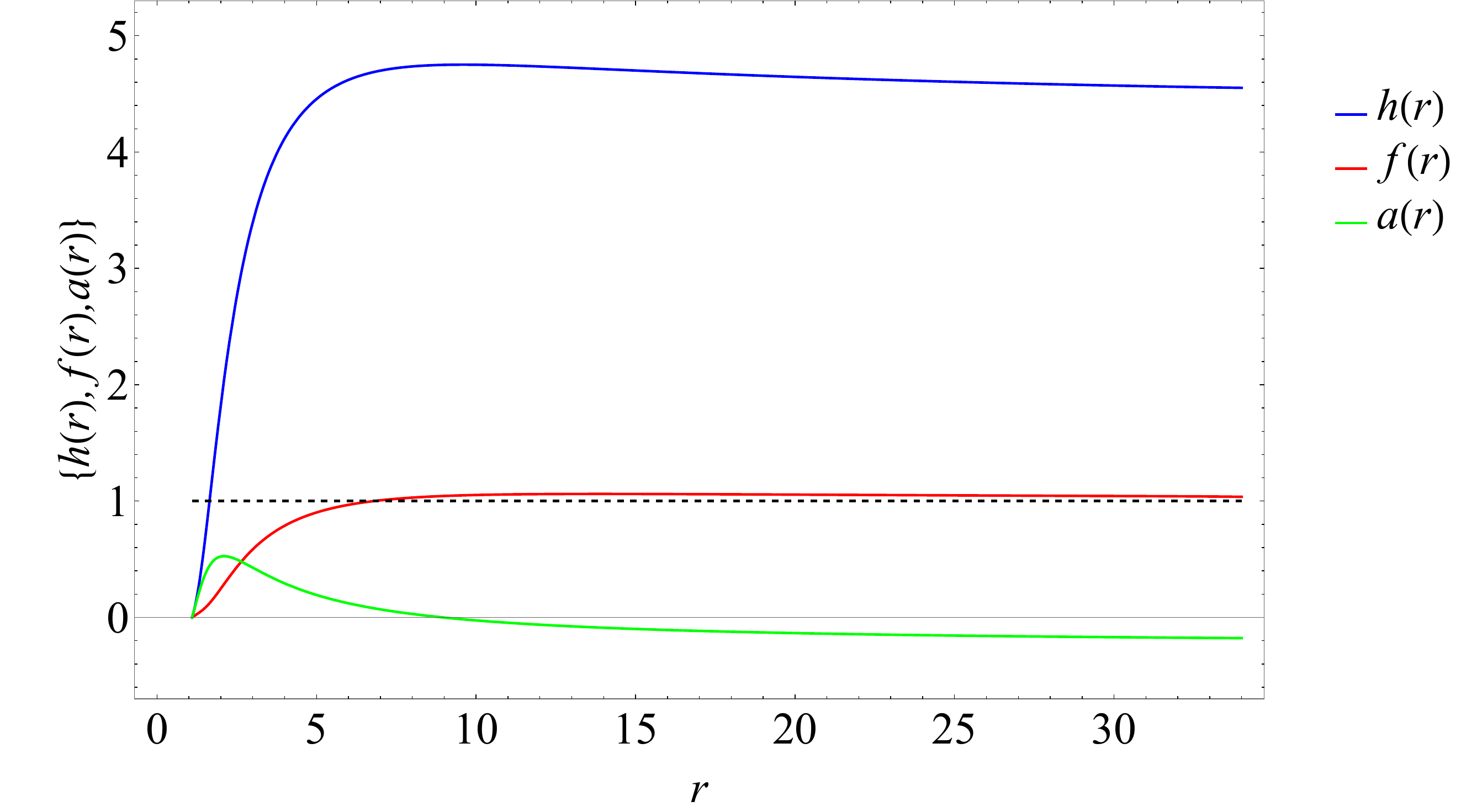}
	\caption{With $a_{1}=1.0$ and $n=1.0$ fixed, the metric functions $h(r), f(r)$, and the electric potential $a(r)$ for the solutions with negative $\delta$ are shown in the plots for $r_{0}=1.1$.}
	\label{Fig3019hfa1}
\end{figure}

With $a_{1}=1.0$ fixed and the NUT parameter increased to $n=1.0$, the general properties of the solutions remain similar to the previous case with $a_{1}=1.0$ and $n=0.7$. In total, there are three branches of numerical solutions. The metric functions and electric potential for $r_0=0.95$ are shown in Figures~\ref{Fig3019hfa} and \ref{Fig3019hfa1}.

\begin{figure}[H]	
	\centering
	\includegraphics[width=0.4\textwidth]{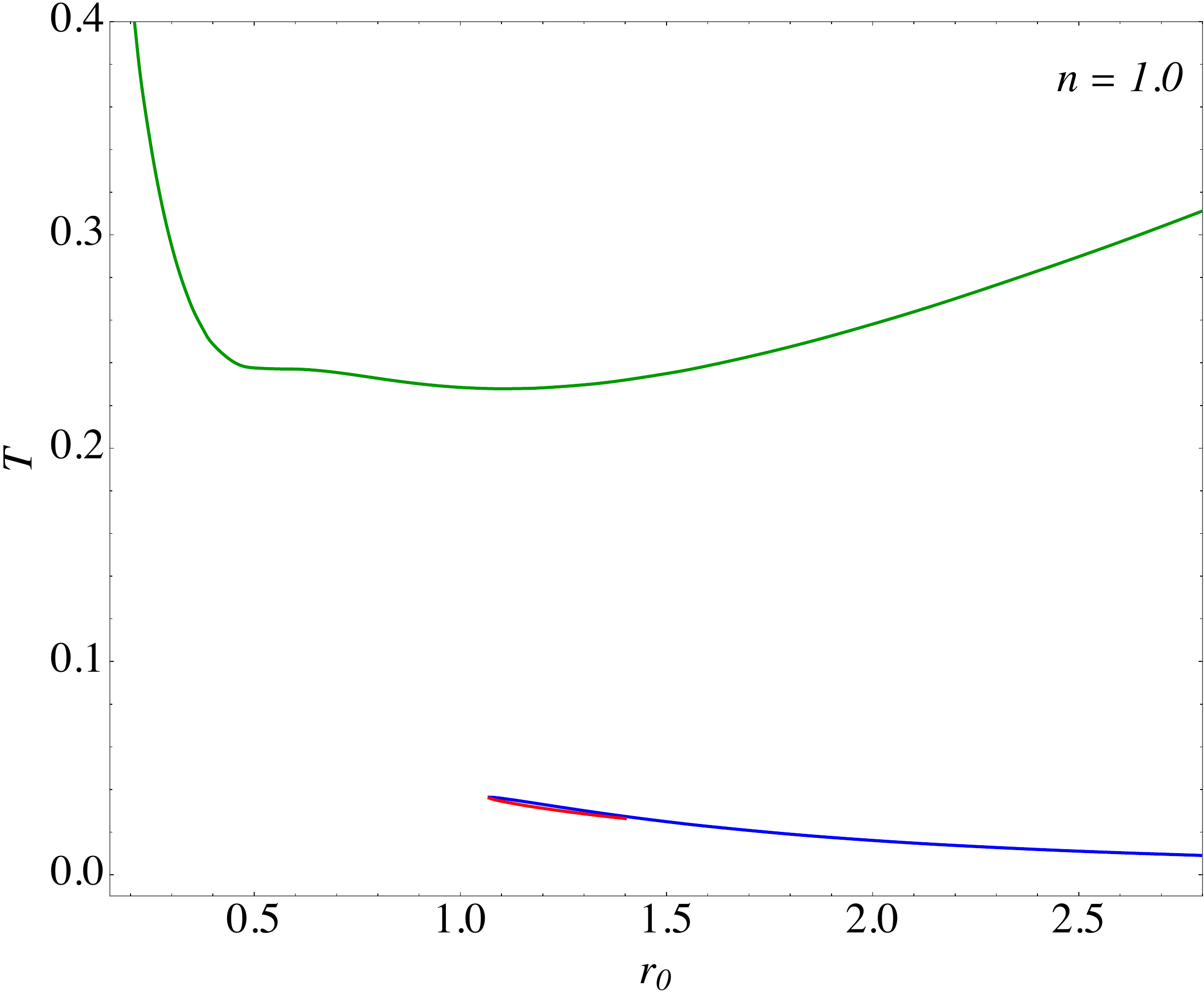}
	\qquad
	\includegraphics[width=0.4\textwidth]{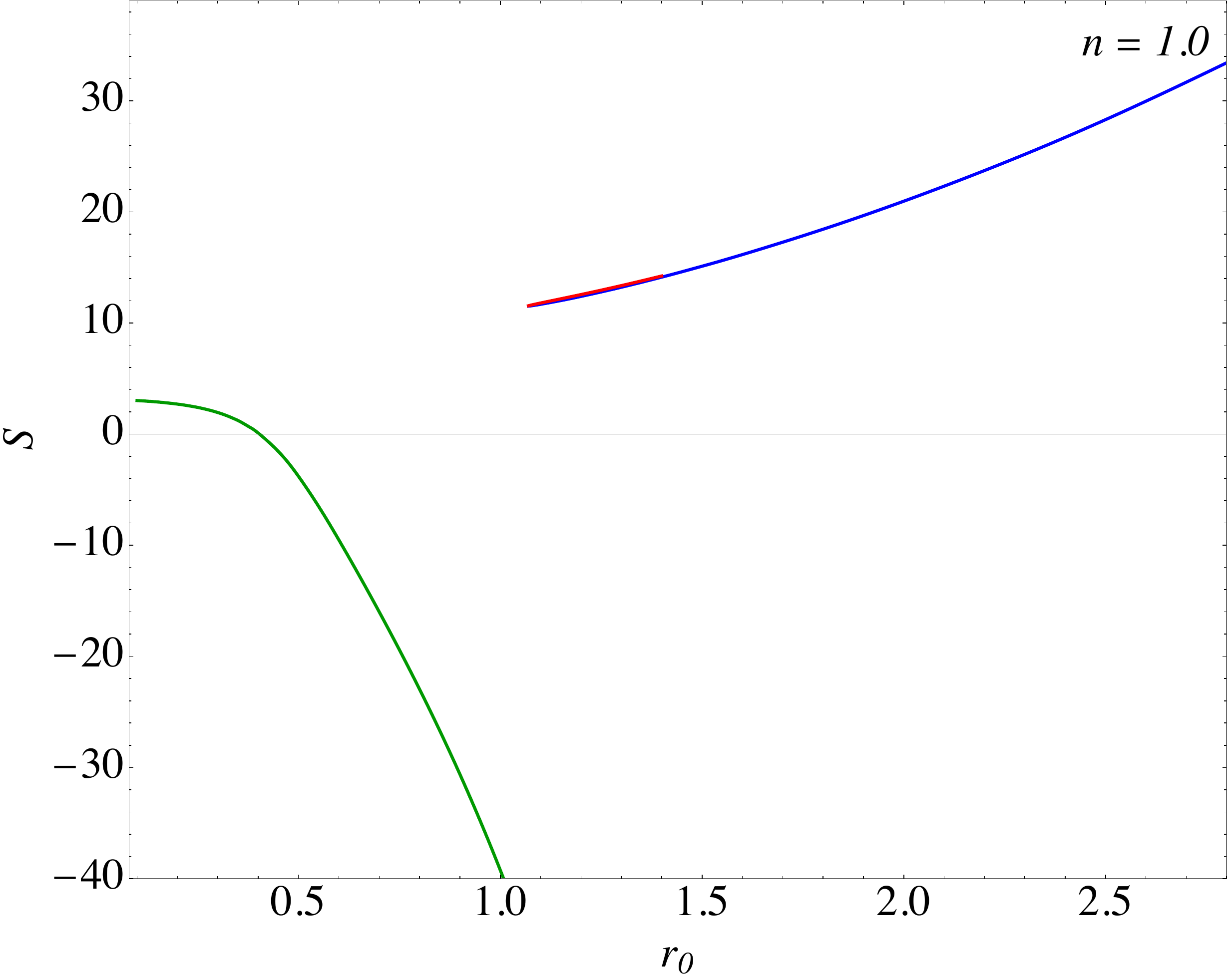}
	\caption{For $a_{1}=1.0$ and $n=1.0$, the temperature and entropy with different $r_{0}$ is shown in the plots. $\delta$ are positive on the green curves and negative on the blue curves and red curves.}
	\label{Fig3021}
\end{figure}

The temperature and entropy as functions of the horizon radius are shown in Figure~\ref{Fig3021}. The three branches of solutions fall into two groups: the solution with positive $\delta$ always has a higher temperature and lower entropy, while the two solutions with negative $\delta$ almost overlap with each other.

\begin{figure}[H]	
	\centering
	\includegraphics[width=0.4\textwidth]{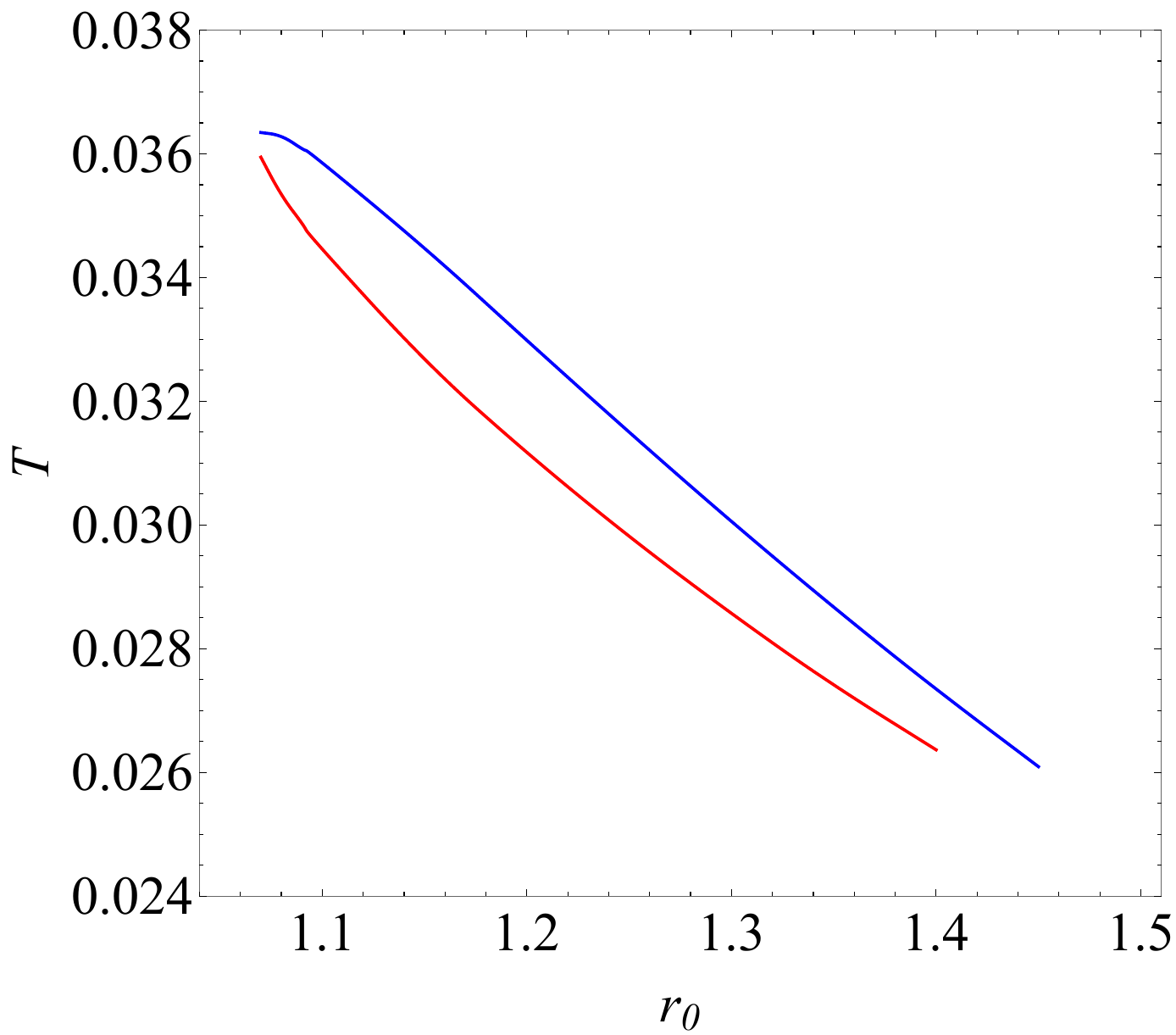}
	\qquad
	\includegraphics[width=0.4\textwidth]{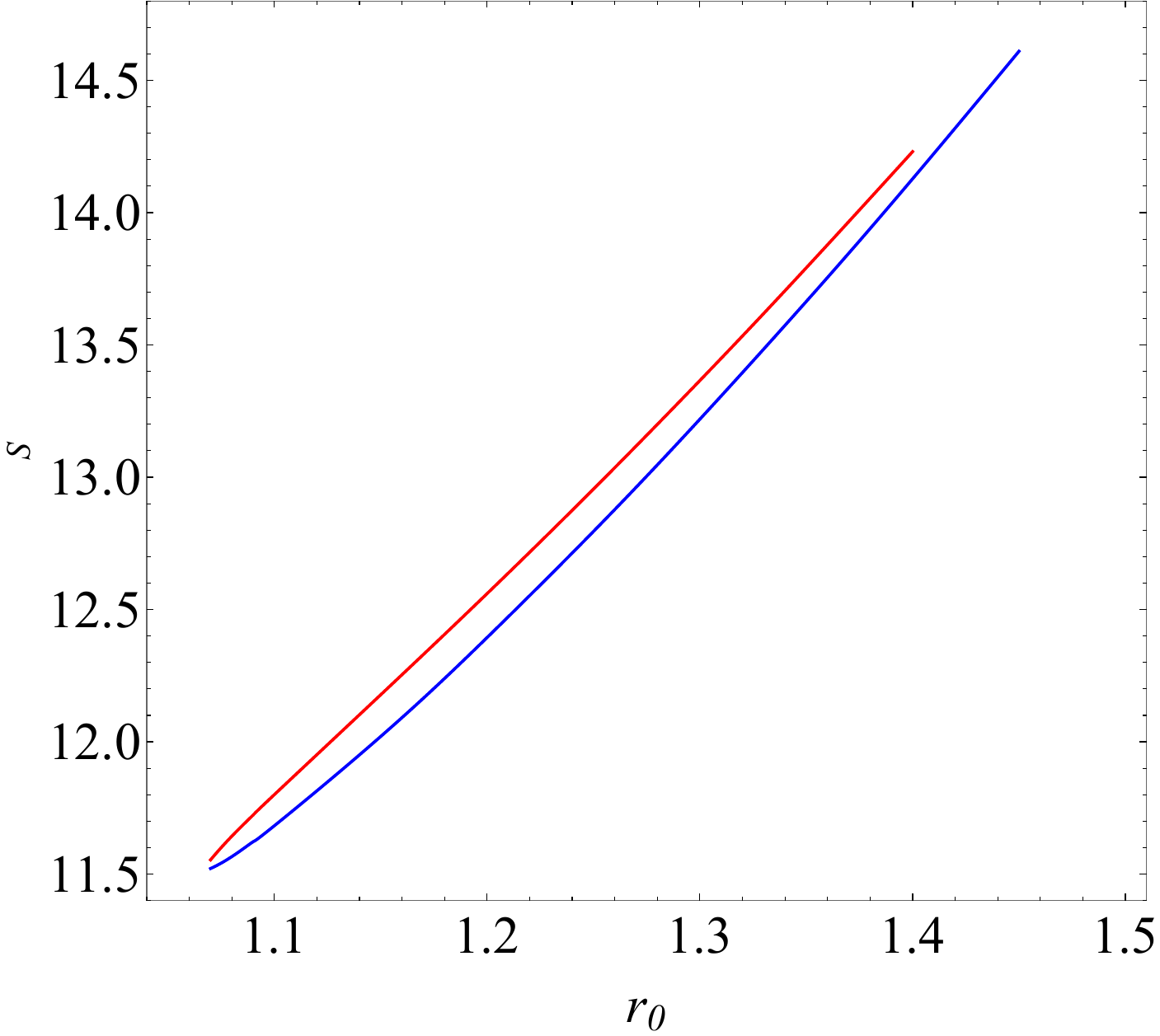}
	\caption{	With $a_{1}=1.0$ and $n=1.0$ fixed, the temperature and entropy for solutions with negative $\delta$ are shown in the plots.}
	\label{Fig3022}
\end{figure}

We enlarge the temperature and entropy curves of the two solutions with negative $\delta$ in Figure~\ref{Fig3022}. As expected, the two branches connect to each other at the minimal horizon radius, where they are smoothly joined.

\begin{figure}[H]	
	\centering
	\includegraphics[width=0.4\textwidth]{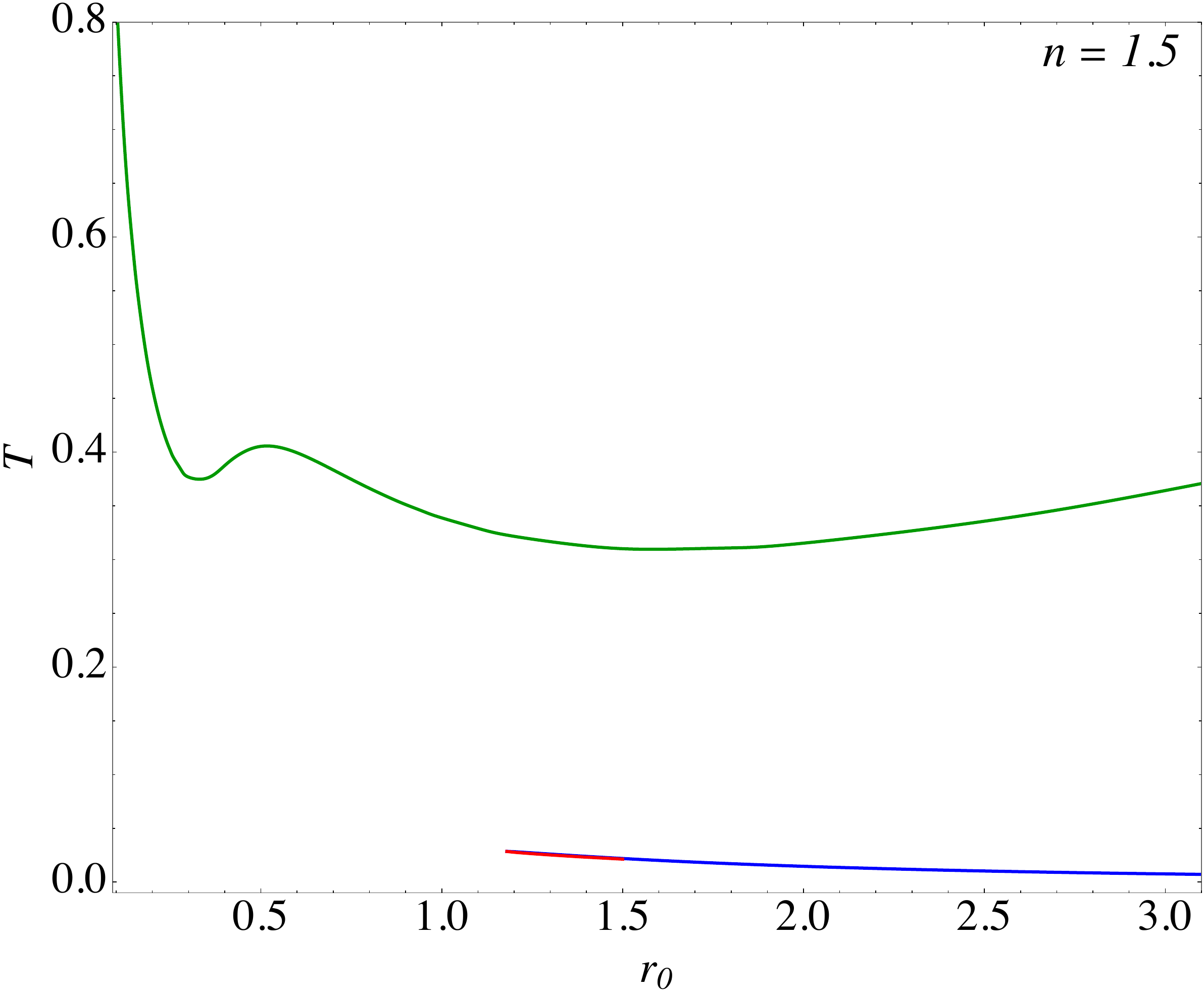}
	\qquad
	\includegraphics[width=0.4\textwidth]{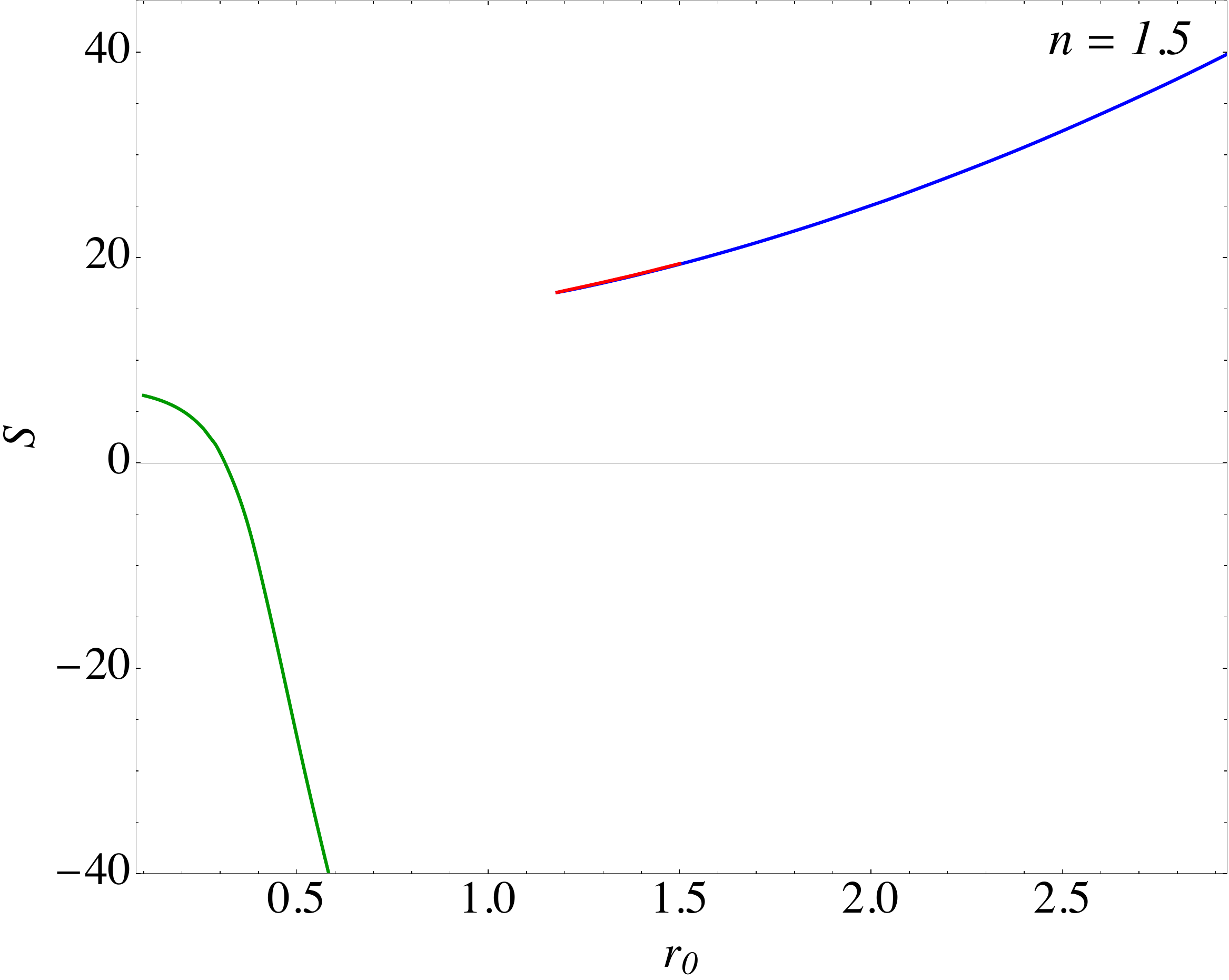}
	\caption{With $a_{1}=1.0$ and $n=1.5$ fixed, the temperature and entropy as functions of $r_{0}$
	are shown in the plots. The parameter $\delta$ is positive on the green curves and negative on the blue and red curves.
	}
	\label{Fig3025}
\end{figure}

With $a_{1}=1.0$ fixed, we further increase the NUT parameter to $n=1.5$ and investigate the resulting numerical solutions. As before, there exist three branches of solutions: one with positive $\delta$ and two with negative $\delta$. The temperature and entropy curves are shown in Figure~\ref{Fig3025}.
The temperature of the charged black hole solution with positive $\delta$ is consistently higher than that of the solutions with negative $\delta$, while its entropy is lower. 

\section{Summary and Discussion}\label{section 4}

We consider the Einstein-Weyl-Maxwell theory in four-dimensional spacetimes and construct new charged Taub-NUT-type black holes within this framework. We carry out detailed study on the entropy and temperature of the newly found black holes. 

 Within a certain parameter range, the theory admits three branches of black hole solutions. These branches can be divided into two groups: the temperature of one group is consistently higher than that of the other. We explore the effects of the charge parameter $a_1$ and the NUT parameter $n$ on these new charged Taub-NUT-like black hole solutions. When the charge parameter vanishes, the solutions reduce to the neutral Taub-NUT-like solutions in Einstein-Weyl theory, where the two groups of solutions intersect at a point~\cite{Chen:2024hsh}. The introduction of the charge parameter causes the two groups of solutions to separate, and they drift further apart as the charge parameter increases. On the other hand, when the NUT parameter vanishes, the solution reduces to static spherical RN-like solutions~\cite{ Li:2025dpo}, and only two branches of solutions exist, rather than three.  Once the NUT parameter is turned on, the lower-temperature solution splits into two branches, which share a common minimal horizon radius where these two branches of solutions smoothly connect to each other. 
 
The thermodynamics of Taub-NUT black holes is a fruitful and still active research area. Considerable work has been carried out, yet challenges remain, particularly regarding the definitions of mass and NUT charge \cite{Hennigar:2019ive,Wu:2019pzr,Wu:2022rmx,Awad:2022jgn,Liu:2022wku}. The inclusion of higher-derivative terms further complicates these issues. The diverse solutions we construct provide a broad platform for investigating the definitions of mass and NUT charge within the framework of higher-derivative gravity theories. And a further extension on this basis is to explore the phase diagram and possible phase transitions of this system. We hope to give out some progress in this direction in the near future.

\acknowledgments
We are grateful to Hong Lu for useful discussion. This work is supported in part by NSFC (National Natural Science Foundation of China) Grants No.~12075166, Tianjin University Graduate Liberal Arts and Sciences Innovation Award Program (2023) No. B1- 2023-005 and Tianjin University Self-Innovation Fund Extreme Basic Research Project Grant No. 2025XJ21-0007.

\bibliographystyle{JHEP}

\bibliography{ref-Einstein-Weyl-Maxwell-1}

\end{document}